\documentclass[journal]{IEEEtran}
\usepackage{cite}

\ifCLASSINFOpdf
  \usepackage[pdftex]{graphicx}
\else
  \usepackage[dvips]{graphicx}
\fi
\usepackage{bm}

\usepackage{amsmath}
\usepackage{amssymb}
\usepackage{algorithmic}

\usepackage{array}
\usepackage{fancyhdr}
\usepackage{lastpage}
\usepackage{layout}
\usepackage{stfloats}

\usepackage{xcolor}
\usepackage{subfigure}

\begin{document}
%
\title{Sinusoidal Parameter Estimation from Signed Measurements via Majorization-Minimization Based RELAX$^{\dag}$}
%
%
%

\author{Jiaying~Ren,
        Tianyi~Zhang,
        Jian~Li,~\IEEEmembership{Fellow,~IEEE},~
        and~Petre~Stoica,~\IEEEmembership{Fellow,~IEEE}
\thanks{$^{\dag}$This work was supported in part by the National Science Foundation (NSF) under Grant No. 1704240 and Grant No. 1708509, and in part by the Swedish Research Council (VR). (Corresponding author: Jian Li)}
\thanks{Jiaying Ren, Tianyi Zhang and Jian Li are with the Department
of Electrical and Computer Engineering, University of Florida, Gainesville,
FL, 32611 USA (e-mail: jiaying.ren@ufl.edu, tianyi.zhang@ufl.edu, li@dsp.ufl.edu.)}
\thanks{P. Stoica is with the Department of Information Technology, Uppsala University,
Uppsala SE-751 05, Sweden (e-mail: ps@it.uu.se).}}

\maketitle

\begin{abstract}
We consider the problem of sinusoidal parameter estimation using signed observations obtained via one-bit sampling with fixed as well as time-varying thresholds. In a previous paper, a relaxation-based algorithm, referred to as 1bRELAX, has been proposed to iteratively maximize the likelihood function. However, the exhaustive search procedure used in each iteration of 1bRELAX is time-consuming. In this paper, we present a majorization-minimization (MM) based 1bRELAX algorithm, referred to as 1bMMRELAX, to enhance the computational efficiency of 1bRELAX. Using the MM technique, 1bMMRELAX maximizes the likelihood function iteratively using simple FFT operations instead of the more computationally intensive search used by 1bRELAX. Both simulated and experimental results are presented to show that 1bMMRELAX can significantly reduce the computational cost of 1bRELAX while maintaining its excellent estimation accuracy.
\end{abstract}


%
\IEEEpeerreviewmaketitle

\section{Introduction}
Signal quantization is a key step in digital signal processing applications which converts an analog signal into a digital signal. The typical case of quantization is to obtain high precision quantized samples, where the quantization error can be modeled as additive noise. However, as the power consumption and cost of an analog-to-digital converter (ADC) grow exponentially with the bit depth $b$, low resolution quantization might be of interest, especially when the sampling rate is high \cite{R99,BTJC05}. Low resolution quantization plays an important role in modern digital signal processing due to its low cost and low power consumption advantages and for allowing for ultra-high sampling rates \cite{R99,BTJC05}. Low resolution sampling has many applications, including spectral sensing for cognitive radios and radars \cite{HACY13,JVH15}, environmental sensing using automotive radars for autonomous driving \cite{JER12,SMDM17,FPAFM17}, and channel estimation for massive multiple-input multiple-output (MIMO) systems \cite{SGMU15,JJR16,C16,CJER17,PJMPP17,KMZA17,AA17,YCGA17,LZLW18,MS18,HA18}.

As an extreme form of low resolution quantization, one-bit sampling (which quantizes the signals using a simple comparator with some reference levels) has attracted much research interests recently \cite{SGMU15,JJR16,C16,CJER17,PJMPP17,KMZA17,AA17,YCGA17,LZLW18,MS18,HA18,FV91,ASFP91,VG98,XY15,JMSM16,SMMJ17,JJ2017,CLJP16,CP06,AFJ10,MKJ16,KYMYZ16,AP00,OA02,PR08,CLJP17,KRR16,CLJP162}. The power consumption of one-bit sampling at a rate of 240 GHz is only about 10 mW, which is much less than the power consumed by a conventional ADC \cite{BTJC05,B13}. The idea of one-bit sampling appeared in early works \cite{W48,J63,WE64,JD66,M68,R70}, and was further theoretically analyzed in \cite{GA94,TV98,AP00,OA02,PR08}. Due to its attractive properties, one-bit sampling has been considered for radar sensing \cite{FV91,ASFP91,VG98,XY15,JMSM16,SMMJ17,JJ2017}, frequency estimation for both temporal and spatial sinusoidal signals \cite{CP06,AFJ10,MKJ16,KYMYZ16,AP00,OA02}, as well as massive MIMO millimeter (mm) wave communications \cite{SGMU15,JJR16,C16,CJER17,PJMPP17,KMZA17,AA17,YCGA17,LZLW18,MS18,HA18}. In most of the previous literature on one-bit sampling, however, the focus was on comparing the signals to zero, which means that signal amplitude information could not be recovered \cite{PR08,KRR16}. In this paper, we consider the problem of sinusoidal parameter estimation using signed measurements obtained via one-bit sampling with fixed as well as time-varying thresholds.

Recently, one-bit sampling using time-varying thresholds has been considered to enable accurate amplitude estimation. In \cite{CLJP16}, the maximum-likelihood (ML) estimator and the corresponding Cramer-Rao bounds (CRBs) are presented and it was shown that one-bit sampling with time-varying thresholds allows for accurate amplitude estimation for either known or unknown noise variances. In \cite{CLJP17}, a relaxation-based method, referred to as 1bRELAX, is proposed for sinusoidal parameter estimation via the maximization of the likelihood function. The 1bRELAX algorithm provides a good estimation performance but still suffers from a high computational burden due to the time-consuming exhaustive searches needed in each iteration. Additionally, sparse methods based on $l_1$ and logarithmic penalty are proposed in \cite{CLJP162}, but the numerical examples in \cite{CLJP17} have demonstrated that 1bRELAX outperforms these sparse methods.

In this paper, a sinusoidal parameter estimation method based on the 1bRELAX framework and majorization-minimization (MM, see, e.g. \cite{HL04,PY04}) approach is proposed. We derive an MM algorithm that can be used to maximize the likelihood function for the signed measurements via simple FFT operations. One key step of our derivation is finding an appropriate majorizing function for the negative log-likelihood of the signed measurements. By using the MM technique, the proposed MM-based 1bRELAX algorithm, referred to as 1bMMRELAX, enhances the computational efficiency of 1bRELAX without sacrificing its excellent estimation accuracy. Note that though the MM approach, which is a type of iterative method for optimization problems \cite{HL04,PY04,J15,MMZJ16,YPD17}, has been widely used in many applications \cite{JMJ06,MJR07,JPD16}, it is not clear how MM can be used to deal with the maximum likelihood problem for the signed measurements. Our main contributions can be summarized as follows:

1) We present a majorization-minimization (MM) based 1bRELAX algorithm, referred to as 1bMMRELAX, for sinusoidal parameter estimation using signed measurments obtained via one-bit sampling  with fixed non-zero or time-varying thresholds.

2) We introduce a proper majorizing function and develop the MM procedure for minimizing the negative log-likelihood function for the signed measurements. The resulting optimization problem at each MM iteration can be interpreted as a sinusoidal parameter estimation problem for infinite precision data, so that the negative log-likelihood function for the signed measurements can be minimized by using the MM approach via simple FFT operations.

3) For the case that the number of sinusoids, i.e., the model order, is unknown, we explain how the proposed algorithm can be used with the one-bit Bayesian information criterion (1bBIC) \cite{LZLP18} to simultaneously estimate the sinusoidal parameters and determine the number of sinusoids.

4) We also extend the proposed algorithm to the complex-valued case and the two-dimensional (2-D) case.

5) Both fixed non-zero and time-varying thresholds are considered to enable accurate amplitude estimation from signed measurements obtained via one-bit sampling.

6) Numerical examples are provided to demonstrate the performance of 1bMMRELAX and 1bBIC for sinusoidal parameter estimation and model order determination. We also compare the performance of 1bMMRELAX with that of several existing methods and with the corresponding CRBs. Additionally, the results obtained by applying 1bMMRELAX with 1bBIC to experimental data for range-Doppler imaging using automotive radar are presented to demonstrate the effectiveness of the proposed method for practical applications.

\textsl{Notation}: We denote vectors and matrices by boldface lower-case and upper-case letters, respectively. $\left(\cdot\right)^T$ and $\left(\cdot\right)^H$ denote the transpose and the conjugate transpose, respectively.              ${\bf R}\in {\mathbb{R}}^{L\times M}$ denotes a real-valued $L\times M$ matrix, and ${\bf R}\in {\mathbb{C}}^{L\times M}$ denotes a complex-valued $L\times M$ matrix. $\mathbb{R}^+$ denotes the positive real numbers. ${\rm Re}\left(x\right)$ and ${\rm Im}\left(x\right)$ denote the real and imaginary parts of $x$, respectively. ${\bf X}=\{x_{n,m}\}$ denotes a matrix whose $\left(n,m\right)$th entry is $x_{n,m}$. ${\rm vec}\left(\cdot\right)$ is the vectorization operator that stacks the columns of a matrix on top of each other. Finally, $j=\sqrt{-1}$.

%
%
%
%

\section{Problem Formulation}
Consider a real-valued one-dimensional (1-D) sinusoidal signal $s_t\left(\bm{\theta}\right)$ \cite{SR05,JP96,AEJS14,CD11}:

\begin{equation}
\begin{split}
s_t\left(\bm{\theta}\right)&=\sum_{k=1}^K A_k {\rm sin}\left(\omega_k t+\phi_k\right)\\
&=\sum_{k=1}^K a_k {\rm sin}\left(\omega_k t\right)+b_k {\rm cos}\left(\omega_k t\right),\\
\label{signal}
\end{split}
\end{equation}
where  $K$ is the number of sinusoids, $A_k\in \mathbb{R}^+$, $\omega_k\in\left[0,\pi\right)$, and $\phi_k\in \left[0,2\pi\right)$ denote the amplitude, frequency, and phase of the $k$th sinusoidal component, respectively, and $t$ denotes the time variable. The unknown sinusoidal parameter vector is denoted by ${\bm{\theta}}=\left[a_1,b_1,\omega_1,\cdots,a_K,b_K,\omega_K \right]^T \in {\mathbb{R}}^{3K}$ with $a_k=A_k\cos \phi_k \in \mathbb{R}$ and $b_k=A_k\sin \phi_k \in \mathbb{R}$.

Suppose that we have $N$ noisy, 1-D real-valued signed measurements, obtained via one-bit sampling with a real-valued time-varying threshold $\{h_n\}_{n=0}^{N-1}$, given by:
\begin{equation}
y_n={\rm{sign}}\left(s_n\left(\bm{\theta}\right)+e_n-h_n\right),
\label{sampling}
\end{equation}
where ${\bf e}=\left[e_0,\cdots,e_{N-1}\right]^T\in \mathbb{R}^{N}$ is the unknown additive noise vector, ${\bf h}=\left[h_0,\cdots,h_{N-1}\right]^T\in \mathbb{R}^{N}$ is the known threshold vector, $n$ is the time index, and ${\rm{sign}\left(\cdot \right)}$ is the sign operator defined as:
\begin{equation}
{\rm{sign}}\left(x\right)= \left\{ \begin{array}{ll}
1 & \textrm{if $x\ge0$},\\
-1 & \textrm{if $x<0$}.\\
\end{array} \right.
\label{sign}
\end{equation}

Under the assumption that the additive noise $e_n$ is i.i.d. Gaussian with zero-mean and unknown variance $\sigma^2$, the likelihood function of the signed measurements is given by \cite{CLJP16,CLJP17}:
\begin{equation}
\begin{split}
&L\left({\bm{\beta}}\right)=\prod_{n=0}^{N-1} \Phi \left(y_n\frac{s_n\left(\bm{\theta}\right)-h_n}{\sigma}\right)\\
&=\prod_{n=0}^{N-1} \Phi \left[y_n\frac{\left(\sum_{k=1}^K a_k{\rm sin}\left(\omega_k n\right)+b_k {\rm cos}\left(\omega_k n\right)\right)-h_n}{\sigma}\right],\\
\label{LF}
\end{split}
\end{equation}
where $\Phi\left(x\right)$ denotes the cumulative distribution function (cdf) of the standard normal distribution and the unknown parameter vector is $\bm{\beta}=\left[\bm{\theta}^T,\sigma\right]^T$. (To simplify the notation, we assume that the sampling period is unity.)

We are interested in estimating the parameter vector $\bm{\beta}$, as well as the order $K$, based on the signed measurement vector ${\bf{y}}=\left[ y_0,y_1,\cdots,y_{N-1}\right]^T \in \left\{-1,+1\right\}^{N}$.

\section{Maximum Likelihood Estimation and 1bRELAX}
\subsection{Maximum Likelihood Estimation}
The maximum likelihood (ML) estimator is a theoretically appealing approach for sinusoidal parameter estimation since it has many desirable properties including consistency, asymptotic efficiency and asymptotic normality. One can obtain the ML estimate of the parameter vector $\bm{\beta}$ by minimizing the negative log-likelihood function \cite{CLJP16}:
\begin{equation}
\begin{split}
&{\bm{\widehat {\widetilde{\beta}}}}={\rm{arg}}\min_{\bm{{\widetilde \beta}}}l \left({\bm{\widetilde {\beta}}}\right)={\rm{arg}}\min_{\bm{\widetilde {\beta}}}\sum_{n=0}^{N-1}\\
&-\!{\rm log}\!\left[\Phi \left(y_n\!\left(\!\left(\sum_{k=1}^K {\widetilde a}_k {\rm sin}\left(\omega_k n\right)\!+\!{\widetilde b}_k {\rm cos}\left(\omega_k n\right)\right)\!-\!\lambda h_n\right)\!\right)\!\right],\\
\end{split}
\label{NLFUN}
\end{equation}
where $\lambda=\frac{1}{\sigma}$, ${\widetilde a}_k=\frac{1}{\sigma}a_k$, ${\widetilde b}_k=\frac{1}{\sigma}b_k$, and the unknown parameter vector is rewritten as $\bm{\widetilde {\beta}}=\left[{\bm{\widetilde {\theta}}}^T,\lambda\right]^T$ with ${\bm{\widetilde {\theta}}}=\left[{\widetilde a}_1,{\widetilde b}_1,\omega_1,\cdots,{\widetilde a}_K,{\widetilde b}_K,\omega_K \right]^T$.

Let $\bm{\omega}=\left[\omega_1,\omega_2,\cdots,\omega_K\right]^T$ be a vector composed of the frequencies of $s_t\left({\bm{\theta}}\right)$. For given $\bm{\omega}$, the above optimization problem is convex in $\{{\widetilde a}_k\}_{k=1}^K$, $\{{\widetilde b}_k\}_{k=1}^K$ and $\lambda$. Therefore, for fixed $\bm{\omega}$, globally optimal methods can be employed to find the minimizing values of $\{{\widetilde a}_k\}_{k=1}^K$, $\{{\widetilde b}_k\}_{k=1}^K$ and $\lambda$ \cite{CLJP16,CLJP17}. With this fact in mind, the ML estimator can be summarized as follows. First, perform a $K$-dimensional search of $\bm{\omega}$ on the feasible space of frequencies $\left[0,\pi\right)^K$, then compute the corresponding optimal $\{{\widehat {\widetilde a}}_k\}_{k=1}^K$, $\{{\widehat {\widetilde b}}_k\}_{k=1}^K$ and $\widehat {\lambda}$ as functions of $\bm{\omega}$. The details on the implementation of the $K$-dimensional frequency search can be found in \cite{CLJP17}.

Note that a direct grid-based implementation of the ML algorithm requires a $K$-dimensional search on $\left[0,\pi\right)^K$. Suppose that there are $L$ points in each dimension. Then, the ($2K+1$)-dimensional convex optimization problem should be solved ${\emph{O}}\left({L^K}\right)$ times. As the number of sinusoids $K$ increases, the search over the high-dimensional frequency space becomes computationally prohibitive and more efficient algorithms must be considered \cite{CLJP16,CLJP17}.

\subsection{1bRELAX and 1bCLEAN}
Inspired by the RELAX algorithm, which is a conceptually and computationally simple method for harmonic retrieval proposed for the infinite precision quantization case \cite{JP96}, the 1bRELAX algorithm was proposed as a relaxation-based approach to maximize the likelihood function \cite{CLJP17}.

\begin{table}
\centering
\caption{1bRELAX}
\begin{tabular}{l}
\hline
1:\;\;\textbf{Input}: Signed measurement vector $\bf y$, the desired or estimated\\
\;\;\;\;\; model order $\widehat K$, and the maximum number of update iterations $I_R$.\\
2:\;\;Assume $K=1$. Obtain $\{{\widehat{\widetilde a}}_1,{\widehat{\widetilde b}}_1,{\widehat{\omega}}_1\}$ and $\lambda$ by solving (\ref{NLFUN})\\
\;\;\;\;\;via the exhaustive search (over ${\omega}_1$).\\
3:\;\;\textbf{for} $K=2:\widehat K$\\
4:\;\;\;\;\;\;$i=0$;\\
5:\;\;\;\;\;\;Repeat:\\
6:\;\;\;\;\;\;\;\;\;\;Obtain $\{{\widehat{\widetilde a}}_K,{\widehat{\widetilde b}}_K,{\widehat{\omega}}_K\}$ by solving (\ref{NLFUN}) via the exhaustive\\
\;\;\;\;\;\;\;\;\;\;\;\;\;search with $\{{{\widetilde a}}_q,{{\widetilde b}}_q,{{\omega}}_q\}_{q=1}^{K-1}$ and $\lambda$ replaced by their \\
\;\;\;\;\;\;\;\;\;\;\;\;\;most recent estimates $\{{\widehat{\widetilde a}}_q,{\widehat{\widetilde b}}_q,{\widehat{\omega}}_q\}_{q=1}^{K-1}$ and $\widehat{\lambda}$;\\
7:\;\;\;\;\;\;\;\;\;\;Redetermine $\{{\widehat{\widetilde a}}_1,{\widehat{\widetilde b}}_1,{\widehat{\omega}}_1\}$ and $\widehat{\lambda}$ by solving (\ref{NLFUN}) via the\\
\;\;\;\;\;\;\;\;\;\;\;\;\;exhaustive search with $\{{{\widetilde a}}_q,{{\widetilde b}}_q,{{\omega}}_q\}_{q=2}^K$ replaced by\\
\;\;\;\;\;\;\;\;\;\;\;\;\;their most recent estimates $\{{\widehat{\widetilde a}}_q,{\widehat{\widetilde b}}_q,{\widehat{\omega}}_q\}_{q=2}^K$;\\
8:\;\;\;\;\;\;\;\;\;\;\textbf{if} $K>2$\\
9:\;\;\;\;\;\;\;\;\;\;\;\;\;\;\textbf{for} $k=2:K-1$\\
10:\;\;\;\;\;\;\;\;\;\;\;\;\;\;\;\;Update $\{{\widehat{\widetilde a}}_k,{\widehat{\widetilde b}}_k,{\widehat{\omega}}_k\}$ by solving (\ref{NLFUN}) via the exhaustive\\
\;\;\;\;\;\;\;\;\;\;\;\;\;\;\;\;\;\;\;\;\;search with $\{{{\widetilde a}}_q,{{\widetilde b}}_q,{{\omega}}_q\}_{q=1,q\neq k}^K$ and $\lambda$ replaced by\\
\;\;\;\;\;\;\;\;\;\;\;\;\;\;\;\;\;\;\;\;\;their most recent estimates $\{{\widehat{\widetilde a}}_q,{\widehat{\widetilde b}}_q,{\widehat{\omega}}_q\}_{q=1,q\neq k}^K$ and $\widehat{\lambda}$.\\
11:\;\;\;\;\;\;\;\;\;\;\;\;\;\textbf{end}\\
12:\;\;\;\;\;\;\;\;\;\;\;\textbf{end}\\
13:\;\;\;\;\;\;\;\;\;\;\;$i=i+1$;\\
14:\;\;\;\;\;Until practical convergence or $i$ reaches the maximum number $I_R$.\\
15:\;\;\textbf{end}\\
16:\;\;\textbf{Output}: $\{{\widehat a}_k\}_{k=1}^{\widehat K}=\{{\widehat {\widetilde {a}}}_k\}_{k=1}^{\widehat K}/{\widehat{\lambda}}$, $\{{\widehat b}_k\}_{k=1}^{\widehat K}=\{{\widehat {\widetilde {b}}}_k\}_{k=1}^{\widehat K}/{\widehat{\lambda}}$,\\
\;\;\;\;\;\;\;$\{\omega_k\}_{k=1}^{\widehat K}$, and $\widehat {\lambda}$\\
\hline
\end{tabular}
\end{table}

The detailed steps of 1bRELAX are depicted in Table \uppercase\expandafter{\romannumeral1}. The exhaustive search procedure in the 1bRELAX iterations is implemented by first performing a coarse search on a uniform grid of $N$ frequencies in $\left[0,\pi\right)$. Then, a finer estimate is obtained by minimizing the objective via the Matlab fmincon function over the interval $\left[\widetilde {\omega}_k-\frac{\pi}{N},\widetilde {\omega}_k+\frac{\pi}{N}\right]$, where $\widetilde {\omega}_k$ is the coarse frequency estimate. The "practical convergence" for each assumed model order, i.e., in each step of the 1bRELAX algorithm is determined by checking the relative change of the objective values $l\left(\widetilde{\bm{\beta}}\right)$ between two consecutive iterations. 1bRELAX estimates the parameters of a new sinusoid based on the sinusoids estimated in the previous steps, and then updates the parameters of each sinusoid iteratively in each step. Since the update procedure is implemented by means of an exhaustive search, 1bRELAX is rather time-consuming.

Note that if we do not update the parameter estimates of each sinusoid iteratively for each model order, i.e., in each step, of 1bRELAX, then the 1bRELAX algorithm becomes a CLEAN-like algorithm (see \cite{H74}, \cite{TA96}, and for an overview, see Section 6.5.7 in \cite{SR05}), referred to as 1bCLEAN hereafter. Specifically, in the $K$th step, 1bCLEAN estimates the parameters of the $K$th sinusoid and updates $\widehat {\lambda}$ by solving (\ref{NLFUN}) via the exhaustive search approach based on the parameters of the $\left(K-1\right)$ sinusoids obtained in the previous steps. Compared with 1bRELAX, the 1bCLEAN algorithm has a lower computational complexity due to avoiding the update of the sinusoidal parameters, but provides less accurate parameter estimates than 1bRELAX, especially when the sinusoidal frequencies are closely spaced.

\textit{Remark} 1: The above discussion assumes that the noise variance is unknown, which is the common case in practical applications. If $\sigma$ is known, the negative log-likelihood function can be minimized more easily since $\lambda$ in (\ref{NLFUN}) is known. The sinusoidal parameter estimation algorithms for known $\sigma$ are similar to those for unknown $\sigma$ and we do not present them to keep this paper concise.

\section{Majorization-Minimization Based 1bRELAX}
In this section, we derive a majorization-minimization (MM) based 1bRELAX algorithm, referred to as 1bMMRELAX, to reduce the computational cost of 1bRELAX. We first derive a computationally efficient MM approach to minimize the negative log-likelihood function $l\left(\widetilde{\bm{\beta}}\right)$ in (\ref{NLFUN}). After that, we introduce the 1bMMRELAX algorithm to estimate the 1-D real-valued sinusoidal parameters from signed measurements. The Bayesian information criterion (BIC) is used with 1bMMRELAX to estimate the number of sinusoids, i.e., the model order, from the one-bit observations. Finally, the proposed algorithm is extended to parameter estimation problems associated with one-bit measurements of both 1-D and 2-D complex-valued sinusoids.
\subsection{Negative Log-likelihood Minimization using MM}
In this subsection, we introduce a majorization-minimization (MM) based method to minimize the negative log-likelihood function $l\left(\widetilde{\bm{\beta}}\right)$ for the signed measurements, i.e., to solve the optimization problem (\ref{NLFUN}).

\subsubsection{Majorization-minimization algorithms}
We start by briefly reviewing the basic idea of MM. MM refers to a type of iterative methods that can transform a hard optimization problem into a sequence of simpler ones \cite{HL04,PY04,J15,MMZJ16,YPD17}. Initialized at a feasible solution ${ \widetilde{{\bm{\beta}}}}^0$, a typical MM algorithm for solving (\ref{NLFUN}) consists of two steps at the $i$th iteration: the majorization step and the minimization step \cite{HL04,PY04,J15,MMZJ16,YPD17}. Specifically, the majorization step is to find a majorizing function $G\left({\widetilde{\bm{\beta}}}|{{\widetilde {\bm{\beta}}}}^{i}\right)$ for $l\left(\widetilde{\bm{\beta}}\right)$, such that:
\begin{equation}
l\left(\widetilde{\bm{\beta}}\right)\le G\left({\widetilde{\bm{\beta}}}|{{\widetilde {\bm{\beta}}}}^{i}\right),
\label{MM1}
\end{equation}
\begin{equation}
l\left({{\widetilde {\bm{\beta}}}}^{i}\right)= G\left({{\widetilde {\bm{\beta}}}}^{i}|{{\widetilde {\bm{\beta}}}}^{i}\right),
\label{MM2}
\end{equation}
where ${{\widetilde {\bm{\beta}}}}^{i}$ is the estimate obtained at the $i$th MM iteration.

Then the second minimization step is to update the parameter vector by solving the minimization problem:
\begin{equation}
{{\widetilde {\bm{\beta}}}}^{i+1} = {\arg}{\min_{\widetilde {\bm{\beta}}}}G\left({\widetilde{\bm{\beta}}}|{{\widetilde {\bm{\beta}}}}^{i}\right).
\label{MM3}
\end{equation}

The objective function $l\left(\widetilde{\bm{\beta}}\right)$ is guaranteed to monotonically decrease since
\begin{equation}
l\left({{\widetilde {\bm{\beta}}}}^{i}\right) =  G\left({{\widetilde {\bm{\beta}}}}^{i}|{{\widetilde {\bm{\beta}}}}^{i}\right)\ge G\left({{\widetilde {\bm{\beta}}}}^{i+1}|{{\widetilde {\bm{\beta}}}}^{i}\right) \ge l\left({{\widetilde {\bm{\beta}}}}^{i+1}\right).
\label{MONO}
\end{equation}
The equality follows from (\ref{MM2}), the first inequality from (\ref{MM3}), and the second inequality from (\ref{MM1}). Note that the monotonicity property of MM still holds under the weaker condition $G\left({{\widetilde {\bm{\beta}}}}^{i+1}|{{\widetilde {\bm{\beta}}}}^{i}\right)\le G\left({{\widetilde {\bm{\beta}}}}^{i}|{{\widetilde {\bm{\beta}}}}^{i}\right)$. Consequently, the minimization of $G\left({\widetilde{\bm{\beta}}}|{{\widetilde {\bm{\beta}}}}^{i}\right)$ is not strictly necessary. We only need to decrease the criterion in (\ref{MM3}) within the MM framework \cite{HL04,PY04}.

\subsubsection{Majorizing the negative log-likelihood}
Finding an appropriate majorizing function is the key step for developing an efficient MM algorithm. To construct a majorizing function for the negative log-likelihood $l\left(\widetilde{\bm{\beta}}\right)$, we proceed as follows: first, define an auxiliary vector ${\bf x}\left({\bm{\widetilde {\beta}}}\right)=\left[x_0\left({\bm{\widetilde {\beta}}}\right),x_1\left({\bm{\widetilde {\beta}}}\right),\cdots,x_{N-1}\left({\bm{\widetilde {\beta}}}\right)\right]^T$ with
\begin{equation}
{x_n}\left({\bm{\widetilde {\beta}}}\right)=y_n\left(s_n\left(\bm{\widetilde {\theta}}\right)-\lambda h_n\right), \quad n=0,\cdots,N-1.
\label{xdefine}
\end{equation}
Then the objective function in (\ref{NLFUN}) can be rewritten as:
\begin{equation}
\begin{split}
l\left({\bm{\widetilde{\beta}}}\right)={\widetilde l}\left(\bf{x}\left({\bm{\widetilde{\beta}}}\right)\right)&=-\sum_{n=0}^{N-1}{\rm{log}}\left(\Phi\left(x_n\left({\bm{\widetilde{\beta}}}\right)\right)\right)\\
&=\sum_{n=0}^{N-1} f\left(x_n\left({\bm{\widetilde{\beta}}}\right)\right),\\
\end{split}
\label{trans}
\end{equation}
where $f\left(x\right)\triangleq -{\rm{log}}\left(\Phi\left(x\right)\right)$. Next, we prove the following results.

 \textit{Lemma 1:} Let $f\left(x\right)\triangleq -{\rm{log}}\left(\Phi\left(x\right)\right)$, $x\in \mathbb{R}$. Then:

1) For all $x\in \mathbb{R}$, $f\left(x\right)$ has a bounded second-order derivative:
\begin{equation}
0< f''\left(x\right)< 1.
\label{BOUND}
\end{equation}

2) For all $x,u\in \mathbb{R}$, the following inequality implied by Taylor's theorem (which is a simple corollary of the mean value theorem) holds:
\begin{equation}
f\left(x\right)\le f\left(u\right)+f'\left(u\right)\left(x-u\right)+\frac{1}{2}\left(x-{u}\right)^2,
\label{Taylore}
\end{equation}
Note that (\ref{Taylore}) becomes an equality for $x=u$.

\begin{figure}
\centering
\includegraphics[width=3in,height=2in]{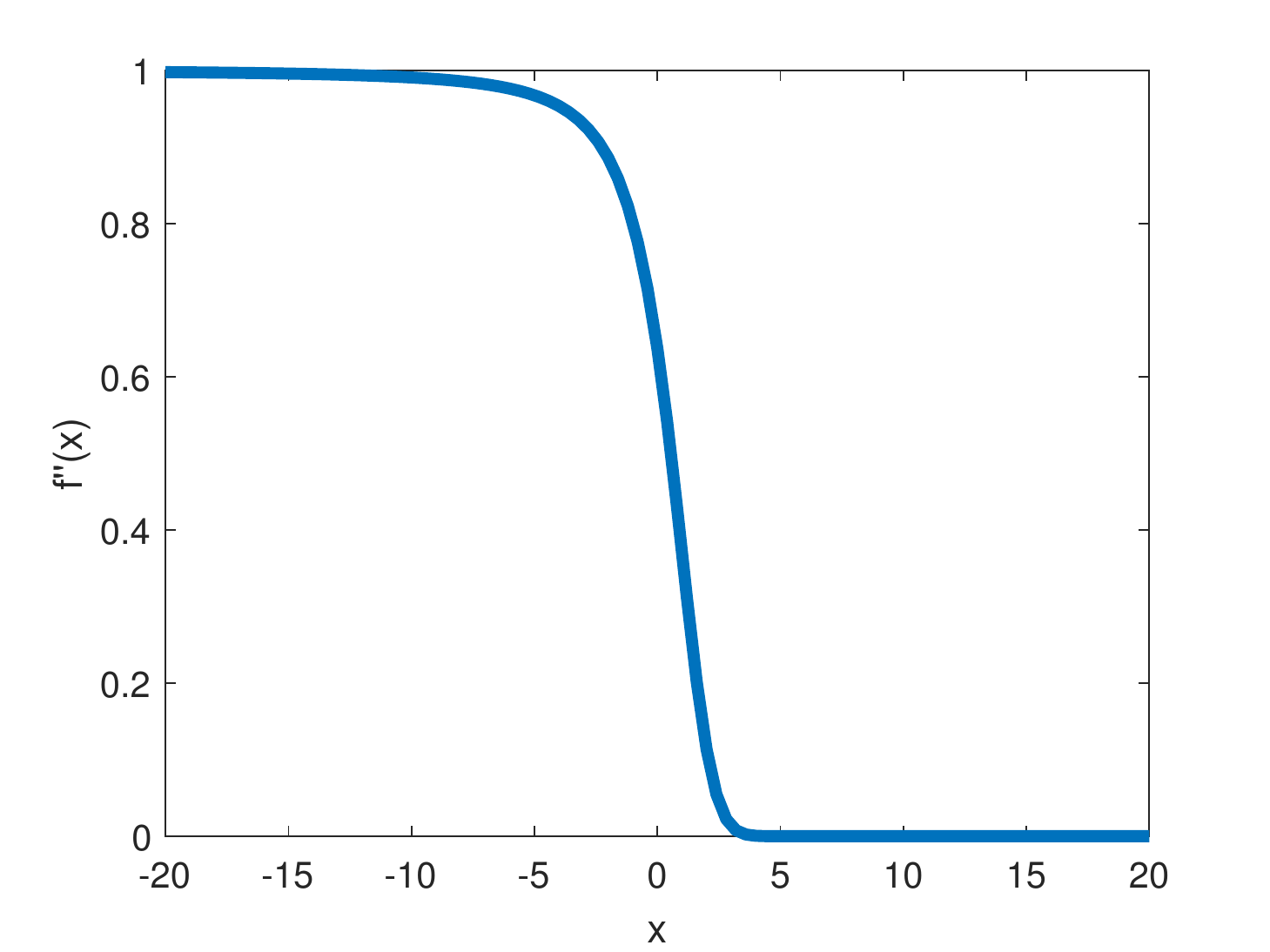}
\caption{The second-order derivative of $f\left(x\right)$.}
\label{Fig.15}
\end{figure}

\textit{Proof:} The first inequality in (\ref{BOUND}) is a direct corollary of a classical result on the inverse Mills ratio (IMR) \cite{S53,I15}. Note that the first derivative of $f\left(x\right)$ can be calculated as:
\begin{equation}
\begin{split}
f'\left(x\right)=&-\frac{\psi\left(x\right)}{\Phi\left(x\right)}=-\frac{{\rm e}^{-\frac{x^2}{2}}}{\int_{-\infty}^{x}{{\rm e}^{-\frac{t^2}{2}}} dt}=-\frac{{\rm e}^{-\frac{\left(-x\right)^2}{2}}}{\int_{-x}^{\infty}{{\rm e}^{-\frac{t^2}{2}}} dt}\\
=&-I_R\left(-x\right),\\
\end{split}
\end{equation}
where $\psi\left(x\right)$ is the standard normal probability density function, and $I_R\left(x\right)$ is known as the IMR. It has been proven that the first derivative of IMR, $I_R'\left(x\right)$, belongs to  the interval $\left(0,1\right)$ for all $x\in \mathbb{R}$ \cite{S53,I15}. Therefore, $0< f''\left(x\right)=I_R'\left(-x\right)< 1$ for all $x\in \mathbb{R}$. Fig. \ref{Fig.15} illustrates this bounded property of $f''\left(x\right)$.

Then, invoking the second-order Taylor theorem \cite{M98} for any $u\in\mathbb{R}$, $f\left(x\right)$ can be written as:
\begin{equation}
f\left(x\right)=f\left(u\right)+f'\left(u\right)\left(x-u\right)+\frac{f''\left(\eta\right)}{2}\left(x-u\right)^2,
\label{Taylor}
\end{equation}
where $\eta$ lies between $x$ and $u$. Since $f''\left(\eta\right)$ is upper bounded as in (\ref{BOUND}), we obtain the upper bound in (\ref{Taylore}) for $f\left(x\right)$.

It follows from the above lemma that given ${\bf x}^i={\bf x}\left(\bm{\widetilde{\beta}}^i\right)$, the estimate obtained at the $i$th MM iteration, we can majorize the objective function of (\ref{NLFUN}) by:
\begin{equation}
\begin{split}
&G\left({\bm{\widetilde{\beta}}}|{\bm{\widetilde {\beta}}}^i\right)={\widetilde G}\left({\bf x}\left({\bm{\widetilde{\beta}}}\right)|{\bf{x}}^i\right)\!=\sum_{n=0}^{N-1}f\left({x}_n^i\right) \\ &+\!f'\left({x}_n^i\right)\left(x_n\left({\bm{\widetilde{\beta}}}\right) \!-\!{x}_n^i\right)+\frac{1}{2}\left(x_n\left({\bm{\widetilde{\beta}}}\right)-{x}_n^i\right)^2.\\
\end{split}
\label{MAX}
\end{equation}

\subsubsection{Update rule and computationally efficient implementation}
With the above observations in mind, the updating formula at the $\left(i+1\right)$th iteration of the MM approach has the following form:
\begin{equation}
\begin{split}
\min_{\bm{\widetilde{\beta}}}& \;{\widetilde G}\left({\bf x}\left({\bm{\widetilde{\beta}}}\right)|{\bf{x}}^i\right)\!=\!\!\min_{\bm{\widetilde{\beta}}} \sum_{n=0}^{N-1}f\left({x}_n^i\right)\!\\
&+\!f'\left({x}_n^i\right)\left(x_n\left({\bm{\widetilde{\beta}}}\right) \!-\!{x}_n^i\right) +\frac{1}{2}\left(x_n\left({\bm{\widetilde{\beta}}}\right)-{x}_n^i\right)^2.\\
\end{split}
\label{UPDATE}
\end{equation}

A simple calculation shows that:
\begin{equation}
\begin{split}
&{\widetilde G}\left({\bf x}\left({\bm{\widetilde{\beta}}}\right)|{\bf {x}}^i\right)\\
&=\sum_{n=0}^{N-1}\frac{1}{2} \left[ x_n^2\left({\bm{\widetilde{\beta}}}\right)-2\left(x_n^i-f'\left({x}_n^i\right)\right)x_n\left({\bm{\widetilde{\beta}}}\right)\right]+{\rm const}\\
&=\sum_{n=0}^{N-1}\frac{1}{2}\left[x_n\left({\bm{\widetilde{\beta}}}\right)-\left(x_n^i-f'\left(x_n^i\right)\right)\right]^2+{\rm const}.\\
\end{split}
\label{SIMPLE}
\end{equation}
It follows from (\ref{SIMPLE}) that the minimization of ${\widetilde G}\left({\bf x}\left({\bm{\widetilde{\beta}}}\right)|{\bf {x}}^i\right)$ in (\ref{UPDATE}) is equivalent to the following problem:
\begin{equation}
\begin{split}
&\min_{{\bm{\widetilde {\beta}}}} \sum_{n=0}^{N-1}\left[x_n\left({\bm{\widetilde{\beta}}}\right)-\left(x_n^i-f'\left(x_n^i\right)\right)\right]^2. \\
\end{split}
\label{SUPDATE}
\end{equation}
Inserting (\ref{xdefine}) into (\ref{SUPDATE}) yields:
\begin{equation}
\begin{split}
&\min_{{\widetilde{\bm{{\beta}}}}} g\left({\widetilde{\bm{{\beta}}}}|{\widetilde{\bm{{\beta}}}}^i\right)\\
&\!=\!\!\min_{{\widetilde{\bm{{\beta}}}}} \sum_{n=0}^{N-1} \left[s_n\left({\widetilde{\bm{\theta}}}\right)-\lambda h_n -y_n\left(x^i_n-f'\left(x^i_n\right)\right)\right]^2 ,\\
\end{split}
\label{BUPDATE}
\end{equation}
We need to solve (\ref{BUPDATE}) at the ($i+1$)th MM iteration. Note that the optimization problem (\ref{BUPDATE}) can be interpreted as a sinusoidal parameter estimation problem for infinite precision data.

\begin{table}
\centering
\caption{MM algorithm for (\ref{NLFUN})}
\begin{tabular}{l}
\hline
1:  \;\;Input: Signed measurement vector $\bf y$, known threshold vector $\bf h$,\\
\;\;\;\; initializations $\bm{\widetilde{\theta}}^0$ and $\lambda^0$, maximum number of MM iterations $I_M$,\\
\;\;\;\; model order $K$, and $i=0$.\\
2:  \;\;Repeat:\\
3:  \;\;\;\;      Update: ${\bm{\widetilde{\beta}}}^i=\left[{\bm{\widetilde{\theta}}}^{iT},{\lambda}^{i}\right]^T$,\\
4:\;\;\;\;  \;${\widetilde z}_n\left(\bm{\widetilde {\beta}}^i\right)=y_n\left(x_n\left(\bm{\widetilde{\beta}}^i\right)-f'\left(x_n\left(\bm{\widetilde {\beta}}^i\right)\right)\right)$, $n=0,\! \cdots,\!N-1$, \\
5: \;\;\;\;  ${\bm{\widetilde{\theta}}}_0^{i+1}={\bm{\widetilde{\theta}}}^i$, $m=0$, $k=K$.\\
6:  \;\;\;\;   Repeat:\\
7:  \;\;\;\;\;\;\;  Update $\lambda_{m+1}^{i+1}={\rm{max}}\left(0,\frac{{\bf h}^T\left[{ {\bf s}}\left({\bm{\widetilde{\theta}}_{m}^{i+1}}\right)-{\widetilde {\bf z}}\left({\bm{\widetilde{\beta}}}^{i}\right)\right]}{{\bf h}^T{\bf h}}\right)$;\\
8:  \;\;\;\;\;\;\;  $v_n^k={\widetilde z}_n\left(\bm{\widetilde {\beta}}^i\right)+\lambda_{m+1}^{i+1}h_n-\sum_{p=1,p\neq k}^K$ \\
\;\;\;\;\;\;\;\; \;\; $\{{\widetilde a}^{i+1}_{p,m}{\rm sin}\left(\omega^{i+1}_{p,m} n\right)+{\widetilde b}^{i+1}_{p,m}{\rm cos}\left(\omega^{i+1}_{p,m} n\right)\}$, $n=0,\! \cdots,\!N-1$;\\
9: \;\;\;\;\;\;\;  Update $\{{\widetilde a}_{k,m+1}^{i+1},{\widetilde b}_{k,m+1}^{i+1},{{\omega}}_{k,m+1}^{i+1}\}$ by using $N_1$-point FFT\\
\;\;\;\;\;\;\;\; \;\; on $\{v_n^k\}_{n=0}^{N-1}$\\
10: \;\;\;\;\;\;\;Refine $\{{\widetilde a}_{k,m+1}^{i+1},{\widetilde b}_{k,m+1}^{i+1},{{\omega}}_{k,m+1}^{i+1}\}$ by using $N_2$-point CZT\\
\;\;\;\;\;\;\;\; \;\; spectral zoom on $\{v_n^k\}_{n=0}^{N-1}$ within the frequency\\
\;\;\;\;\;\;\;\; \;\; interval $\left[{ {\omega}}_{k,m+1}^{i+1}-\frac{2\pi}{N_1},{ {\omega}}_{k,m+1}^{i+1}+\frac{2\pi}{N_1}\right]$\\
11:  \;\;\;\;\;\;\;Redetermine $\{{\widetilde a}_{k,m+1}^{i+1},{\widetilde b}_{k,m+1}^{i+1},{{\omega}}_{k,m+1}^{i+1}\}$ by using $N_2$-point\\
\;\;\;\;\;\;\;\; \;\; CZT spectral zoom on $\{v_n^k\}_{n=0}^{N-1}$ within the \\
\;\;\;\;\;\;\;\; \;\; frequency interval $\left[{ {\omega}}_{k,m+1}^{i+1}-\frac{4\pi}{N_{1}N_2},{ {\omega}}_{k,m+1}^{i+1}+\frac{4\pi}{N_{1}N_{2}}\right]$\\
12: \;\;\;\;\;\;\; $\bm{\widetilde {\theta}}^{i+1}_{m+1}=\left[{\widetilde a}_{1,m}^{i+1},{\widetilde b}_{1,m}^{i+1},{{\omega}}_{1,m}^{i+1},\cdots,{\widetilde a}_{k,m+1}^{i+1},{\widetilde b}_{k,m+1}^{i+1},{{\omega}}_{k,m+1}^{i+1},\right.$\\
\;\;\;\;\;\;\;\; \;\;\;$\left.\cdots,{\widetilde a}_{K,m}^{i+1},{\widetilde b}_{K,m}^{i+1},{{\omega}}_{K,m}^{i+1}\right]$;\\
13: \;\;\;\;\;\;\; $k={\rm mod}\left(k,K\right)+1$, where mod($\cdot$) denotes the modulo \\
\;\;\;\;\;\;\;\; \;\;\;operation; \\
14: \;\;\;\;\;\;\; $m=m+1$;\\
15\;\;\;\;   Until practical convergence.\\
16:  \;\;\;$i=i+1$;\\
17: Until practical convergence or $i$ reaches the maximum number $I_M$.\\
18: Output: $\bm{\widehat{\widetilde {\theta}}}$, $\widehat{\lambda}$.\\
\hline
\end{tabular}
\end{table}

Let ${\bf {\widetilde z}}\left(\bm{\widetilde {\beta}}\right)=\left[{\widetilde z}_0\left(\bm{\widetilde{\beta}}\right),\cdots,{\widetilde z}_{N-1}\left(\bm{\widetilde {\beta}}\right)\right]^T$ with
\begin{equation}
 {\widetilde z}_n\left(\bm{\widetilde {\beta}}\right)\!=\!y_n\left(x_n\left(\bm{\widetilde{\beta}}\right)-f'\left(x_n\left(\bm{\widetilde {\beta}}\right)\right)\right),\quad n=0,\cdots,N-1.
\end{equation}
Using this notation, we can write the optimization problem in (\ref{BUPDATE}) as follows:
\begin{equation}
\min_{\bm{\widetilde{\theta}},\lambda} g \left({\bm{\widetilde{\theta}},\lambda}|\bm{\widetilde{\beta}}^i\right)=\min_{\bm{\widetilde{\theta}},\lambda} \sum_{n=0}^{N-1}\left[{ s}_n\left(\bm{\widetilde {\theta}}\right)\!-\! \lambda h_n-{\widetilde z}_n\left(\bm{\widetilde {\beta}}^i\right)\right]^2.
\label{UNKNOWN}
\end{equation}
The minimization of the objective function in (\ref{UNKNOWN}) can be conveniently achieved by a cyclic algorithm \cite{PY04} that alternates between the minimization of $g\left({\bm{\widetilde{\theta}},\lambda}|\bm{\widetilde{\beta}}^i\right)$ with respect to $\lambda$ for fixed $\bm{\widetilde {\theta}}$ and the minimization of $g\left({\bm{\widetilde{\theta}},\lambda}|\bm{\widetilde{\beta}}^i\right)$ with respect to $\bm{\widetilde {\theta}}$ for given $\lambda$. The solution of the first step can be easily obtained in closed-form as follows:
\begin{equation}
\lambda_{m+1}^{i+1}={\rm{max}}\left(0,\frac{{\bf h}^T\left[{ {\bf s}}\left({\bm{\widetilde{\theta}}_m^{i+1}}\right)-{\widetilde {\bf z}}\left({\bm{\widetilde{\beta}}}^{i}\right)\right]}{{\bf h}^T{\bf h}}\right),
\label{LAMBDA}
\end{equation}
where ${ {\bf s}}\left({\bm{\widetilde{\theta}}_m^{i+1}}\right)=\left[{ {s}}_0\left({\bm{\widetilde{\theta}}_m^{i+1}}\right),\cdots,{ { s}}_{N-1}\left({\bm{\widetilde{\theta}}_m^{i+1}}\right)\right]^T\in \mathbb{R}^{N}$, and the index $m$ denotes the iteration number in the cyclic minimization performed at the $i$th MM iteration. Note that the monotonicity property of the cyclic algorithm holds if
\begin{equation}
\begin{split}
g\left({\bm{\widetilde{\theta}}_{m}^{i+1}},\lambda_{m}^{i+1}|\bm{\widetilde{\beta}}^i\right)& \ge g\left({\bm{\widetilde{\theta}}_{m}^{i+1}},\lambda_{m+1}^{i+1}|\bm{\widetilde{\beta}}^i\right) \\
&\ge g\left({\bm{\widetilde{\theta}}_{m+1}^{i+1}},\lambda_{m+1}^{i+1}|\bm{\widetilde{\beta}}^i\right).\\
\end{split}
\label{MONO2}
\end{equation}
The first inequality in (\ref{MONO2}) follows from the minimization of $g\left({\bm{\widetilde{\theta}}_{m}^{i+1}},\lambda|\bm{\widetilde{\beta}}^i\right)$ with respect to $\lambda$, and the second inequality can be ensured by decreasing $g\left({\bm{\widetilde{\theta}},\lambda_{m+1}^{i+1}}|\bm{\widetilde{\beta}}^{i}\right)$ with respect to $\bm{\widetilde{\theta}}$. Viewing $\{\lambda_{m+1}^{i+1} h_n+{\widetilde z}_n\left(\bm{\widetilde{\beta}}^i\right)\}_{n=0}^{N-1}$ as the input data, $g\left({\bm{\widetilde{\theta}},\lambda_{m+1}^{i+1}}|\bm{\widetilde{\beta}}^{i}\right)$ can be decreased efficiently by an $N_1$-point zero-padded FFT and subsequent spectral zoom operations, i.e., $N_2$-point chirp z-transforms (CZTs) \cite{LRC69,G06}, iteratively, for one sinusoid at a time \cite{JP96}. Note that the CZT can be implemented by simple FFT and IFFT operations \cite{LRC69,G06}. $N_1$ is selected as the smallest power of 2 greater than or equal to $N$, and $N_2$ is chosen as a proper number so that the FFTs and IFFTs performed in CZT are for powers of 2 operations. Additionally, since there exists a simple closed-form solution for ${\lambda}$, we redetermine ${\lambda}$ after updating each sinusoid.

The MM approach for solving the optimization problem in (\ref{NLFUN}) is summarized in Table \uppercase\expandafter{\romannumeral2}. The "practical convergence" of the cyclic algorithm within each MM iteration is determined by checking the relative change of the objective values of (\ref{UNKNOWN}) between two consecutive iterations, and the "practical convergence" of the MM iterations is obtained when the relative change of the objective values of (\ref{NLFUN}) between two consecutive iterations is below a small threshold.

\subsubsection{Discussions}
Now we consider the convergence of the MM algorithm. The technical conditions in \cite{J15} (see Proposition 2.6 in \cite{J15}) guaranteeing that the proposed MM algorithm converges to a stationary point of the function in (\ref{NLFUN}) can be shown to hold under some mild assumptions, by means of straightforward calculations (which we omit in the interest of brevity).

Then, we briefly discuss the computational complexity of the MM algorithm for maximizing the likelihood function for signed measurements. Since the MM algorithm is implemented by means of simple FFT operations, the per-iteration computational complexity of the MM algorithm is $O\left(N \log _2 N\right)$.

\subsection{1bMMRELAX}
We now present the 1bMMRELAX algorithm for efficiently estimating the 1-D real-valued sinusoidal parameters from signed measurements. The basic idea is to speed up the 1bRELAX algorithm through the use of the MM technique. The 1bMMRELAX algorithm is obtained by replacing the update procedure of 1bRELAX (see Step 7-14 of Table \uppercase\expandafter{\romannumeral1}) by the MM algorithm proposed in the above subsection. Specifically, the 1bMMRELAX algorithm begins by assuming $K=1$. In the $K$th step, we first use the $N$-point exhaustive coarse search (in the frequency domain) to get the initial parameters of the $K$th sinusoid making use of the $\left(K-1\right)$ sinusoids obtained in the previous $\left(K-1\right)$ steps. Next, the algorithm refines the parameter estimates of the $K$ sinusoids, starting with the $K$th sinusoid, by using the MM technique to maximize the likelihood function. The MM algorithm is initialized with $\bm{\widetilde {\theta}}^0=\left[{\widehat {\widetilde a}}_1,{\widehat{\widetilde b}}_1,{\widehat{\omega}}_1,\cdots, {\widehat {\widetilde a}}_K,{\widehat{\widetilde b}}_K,{\widehat{\omega}}_K\right]$ and $\lambda^0=\widehat{\lambda}$ provided by the previous $(K-1)$ steps and the coarse search of the $K$th step. We then increase the model order by one in the next step and estimate the signal parameters similarly. The algorithm proceeds until the desired or estimated model order is reached.

We conclude this subsection with the following comments:

1) A good initial estimate is of significant importance to the MM approach since the objective function in (\ref{NLFUN}) is not convex in $\{\omega_k\}_{k=1}^K$. With the exhaustive coarse search used for initialization at each model order, i.e., in each step of 1bMMRELAX, good initial estimates $\{{\widehat{\widetilde a}}_k,{\widehat{\widetilde b}}_k,{\widehat {\omega}}_k\}_{k=1}^K$ are provided to the MM approach.

2) The 1bMMRELAX algorithm refines the parameter estimates via the MM approach based on simple FFT operations with a low computational cost. Therefore, the main computational burden of 1bMMRELAX is due to the exhaustive coarse searches used for initializations. Consequently, 1bMMRELAX has a computational complexity of $O\left(KN^2\right)$, similar to that of 1bCLEAN. Compared to the $O\left(CKN^2+CK^2N^2\right)$ flops required by 1bRELAX, where $C$ is the number of iterations required to achieve practical convergence at each model order, the proposed 1bMMRELAX algorithm has a significantly lower computational complexity, especially when the number of sinusoids is large.

\subsection{Determining the Number of Sinusoids}
To determine the number of sinusoids, i.e., the model order $K$, from signed measurements, we will use the one-bit Bayesian information criterion (1bBIC) with 1bMMRELAX. Suppose that  ${\widehat {\bm {\beta}}}_{\breve{K}}$ is the vector of parameter estimates obtained by the 1bMMRELAX algorithm for an assumed model order ${\breve{K}}$; then the 1bBIC cost function is given by \cite{LZLP18}:
\begin{equation}
\begin{split}
{\rm 1bBIC}\left({\breve K}\right)=&-2\sum_{n=0}^{N-1}{\rm log}\left(\Phi \left(y_n\frac{s_n\left(\bm{\widehat{\theta}}_{\breve{K}}\right)-h_n}{\widehat{\sigma}}\right)\right)\\
&+5\breve{K}{\rm log}N.\\
\end{split}
\label{BIC}
\end{equation}
The estimate $\widehat{K}$ of $K$ is determined as the integer that minimizes the 1bBIC cost function with respect to the assumed number of sinusoids $\breve{K}$.

\subsection{Extension to the Case of 1-D Complex-Valued Sinusoids}
The proposed 1bMMRELAX algorithm for estimating 1-D real-valued sinusoidal parameters from 1-D real-valued signed measurements can be modified to deal with the 1-D complex-valued harmonic retrieval problem. The data model for the 1-D complex-valued case can be written as:
\begin{equation}
\begin{split}
s_{{\rm }t}\left(\bm{\theta}\right)&=\sum_{k=1}^K A_k {\rm e}^{j\left(\omega_k t+\phi_k\right)} \\
&=\sum_{k=1}^K \left[a_k {\rm cos}\left(\omega_k t\right)-b_k {\rm sin}\left(\omega_k t\right)\right]\\
&+j\; \left[a_k {\rm sin}\left(\omega_k t\right)+b_k {\rm cos}\left(\omega_k t\right)\right],\\
\end{split}
\end{equation}
where $A_k {\rm e}^{j\phi_k}=a_k+jb_k$, and the unknown sinusoidal parameter vector is ${\bm{\theta}}=\left[a_1,b_1,\omega_1,\cdots,\right.$ $\left.a_K,b_K,\omega_K \right]^T \in {\mathbb{R}}^{3K}$. For notational simplicity, we use the same symbol $s_{{\rm }t}\left(\bm{\theta}\right)$ to denote the signal, as in (\ref{signal}), despite the fact that it is a complex-valued quantity in this section (the same is true for other symbols used in the equations that follow).

By comparing the noisy signal samples to a complex-valued reference threshold vector ${\bf h}=\left[h_0,\cdots,h_{N-1}\right]^T \in{\mathbb{C}}^{N}$, we obtain $N$ complex-valued signed measurements ${{\bf y}}=\left[y_{{\rm }0},\cdots,\right.$ $\left.y_{{\rm }N-1}\right]^T\in{\{1+j,1-j,-1+j,-1-j\}}^{N}$:
\begin{equation}
y_{{\rm }n}={\rm signc} \left(s_{{\rm }n}\left(\bm{\theta}\right)+e_{{\rm }n}-h_{{\rm }n}\right),
\end{equation}
where ${{\bf e}_{{\rm }}}=\left[e_{{\rm }0},\cdots,e_{{\rm }N-1}\right]^T \in{\mathbb{C}}^{N}$ is the unknown additive noise vector, and ${\rm signc}\left(\cdot\right)$ is the sign-like operator for complex-valued data:
\begin{equation}
{\rm signc}\left(x\right)={\rm sign}\left({\rm Re}\left(x\right)\right)+j{\rm sign}\left({\rm Im}\left(x\right)\right).
\label{signc}
\end{equation}

Assume that the additive noise $e_{n}$ is i.i.d. circularly symmetric complex-valued Gaussian with zero-mean and unknown variance $\sigma^2$, i.e., ${\rm Re}\left(e_{n}\right)\sim N\left(0,\frac{\sigma^2}{2}\right)$ and ${\rm Im}\left(e_{n}\right)\sim N\left(0,\frac{\sigma^2}{2}\right)$. Then, the likelihood function for the complex-valued signed measurements is given by:
\begin{equation}
\begin{split}
L\left({\bm{\beta}}\right)=\prod_{n=0}^{N-1} & \Phi \left({\rm Re}\left(y_{n}\right)\frac{{\rm Re}\left(s_{n}\left(\bm{\theta}\right)\right)-{\rm Re}\left(h_{n}\right)}{\sigma/\sqrt{2}}\right)\\
&\Phi \left({\rm Im}\left(y_{n}\right)\frac{{\rm Im}\left(s_{n}\left(\bm{\theta}\right)\right)-{\rm Im}\left(h_{n}\right)}{\sigma/\sqrt{2}}\right),\\
\label{LF2}
\end{split}
\end{equation}
where the unknown parameter vector is $\bm{\beta}=\left[\bm{\theta}^T,\sigma\right]^T$ or, equivalently, $\bm{\widetilde {\beta}}=\left[{\bm{\widetilde {\theta}}}^T,\lambda\right]^T$ with $\lambda=\frac{1}{\sigma/\sqrt{2}}$.

For the 1-D complex-valued signed measurements, the $\left(i+1\right)$th MM iteration becomes
\begin{equation}
\min_{\bm{\widetilde{\theta}},\lambda} g\left({\bm{\widetilde{\theta}},\lambda}|\bm{\widetilde{\beta}}^i\right)=\min_{\bm{\widetilde{\theta}},\lambda} \sum_{n=0}^{N-1}\left|{ s}_{n}\left(\bm{\widetilde {\theta}}\right)-\lambda h_{n}-{\widetilde z}_{n}\left(\bm{\widetilde {\beta}}^i\right)\right|^2,
\label{UNKNOWN2}
\end{equation}
where
\begin{equation}
\begin{split}
& {\widetilde z}_{n}\left(\bm{\widetilde {\beta}}\right)\\
& =\!{\rm Re}\left(y_{n}\right)\left({\rm Re}\left(x_{{}n}\left(\bm{\widetilde{\beta}}\right)\right)-f'\left({\rm Re}\left(x_{{}n}\left(\bm{\widetilde {\beta}}\right)\right)\right)\right)\\
& +j{\rm Im}\left(y_{{}n}\right)\left({\rm Im}\left(x_{{}n}\left(\bm{\widetilde{\beta}}\right)\right)-f'\left({\rm Im}\left(x_{{}n}\left(\bm{\widetilde {\beta}}\right)\right)\right)\right),\\
\label{zupdate}
\end{split}
\end{equation}
with
\begin{equation}
\begin{split}
x_{{}n}\left(\bm{\widetilde {\beta}}\right)&={\rm Re}\left(y_{{}n}\right)\left({\rm Re}\left(s_{{}n}\left(\bm{\widetilde {\beta}}\right)-\lambda h_{{}n}\right)\right)\\
&+j{\rm Im}\left(y_{{}n}\right)\left({\rm Im}\left(s_{{}n}\left(\bm{\widetilde {\beta}}\right)-\lambda h_{{}n}\right)\right).\\
\end{split}
\end{equation}
Similarly to the real-valued case, (\ref{UNKNOWN2}) can be conveniently solved by the same type of cyclic algorithm. At the $\left(m+1\right)$th cyclic iteration performed within the $\left(i+1\right)$th MM iteration, minimizing (\ref{UNKNOWN2}) for fixed $\widetilde {\bm{\theta}}_{m}^{i+1}$ yields:
\begin{equation}
\lambda_{m+1}^{i+1}={\rm{max}}\left(0,\frac{{\rm Re}\left\{{\bf h}^H\left[{ {\bf s}}\left({\bm{\widetilde{\theta}}_m^{i+1}}\right)-{\widetilde {\bf z}}\left({\bm{\widetilde{\beta}}}^{i}\right)\right]\right\}}{{\bf h}^H{\bf h}}\right).
\label{LAMBDA2}
\end{equation}
where ${ {\bf s}}\left({\bm{\widetilde{\theta}}_m^{i+1}}\right)=\left[{ {s}}_{ 0}\left({\bm{\widetilde{\theta}}_m^{i+1}}\right),\cdots,{ { s}}_{ N-1}\left({\bm{\widetilde{\theta}}_m^{i+1}}\right)\right]^T\in \mathbb{C}^{N}$, and ${ {\widetilde{\bf z}}}\left({\bm{\widetilde{\beta}}^{i}}\right)=\left[{ {\widetilde z}}_{ 0}\left({\bm{\widetilde{\beta}}^{i}}\right),\cdots,\right.$ $\left.{ { \widetilde z}}_{ N-1}\left({\bm{\widetilde{\beta}}^{i}}\right)\right]^T\in \mathbb{C}^{N}$. Next, for given $\lambda_{m+1}^{i+1}$, the function $g\left({{\bm{\widetilde{\theta}}},\lambda_{m+1}^{i+1}}|\bm{\widetilde{\beta}}^i\right)$ can be decreased efficiently by using FFTs and CZT spectral zooms as discussed in Section IV.A. We omit further details of 1bMMRELAX for 1-D complex-valued signed measurements since the modifications are straightforward.

Note that for 1D complex sinusoids, the 1bBIC function becomes \cite{LZLP18}:
\begin{equation}
\begin{split}
&{\rm 1bBIC}\left({\breve K}\right)\\
&=-2\sum_{n=0}^{N-1}{\rm log}\left(\Phi \left({\rm Re}\left(y_{{}n}\right)\frac{{\rm Re}\left(s_{n}\left(\bm{\widehat{\theta}}\right)\right)-{\rm Re}\left(h_{n}\right)}{\widehat{\sigma}/\sqrt{2}}\right)\right)\\
&-2\sum_{n=0}^{N-1}{\rm log}\left(\Phi \left({\rm Im}\left(y_{n}\right)\frac{{\rm Im }\left(s_{n}\left(\bm{\widehat{\theta}}\right)\right)-{\rm Im}\left(h_{n}\right)}{\widehat{\sigma}/\sqrt{2}}\right)\right)\\
&+5\breve{K}{\rm log}N.\\
\end{split}
\label{BICc}
\end{equation}

\subsection{Extension to the Case of 2-D Complex-Valued Sinusoids}
The 1bMMRELAX algorithm can also be straightforwardly extended to the case of 2-D complex-valued sinusoids. The 2-D complex-valued sinusoidal signal can be expressed as:
\begin{equation}
\begin{split}
&s_{t_1,t_2}\left(\bm{\theta}_{\rm 2D}\right)=\sum_{k=1}^K A_k {\rm e}^{j\left(\omega_{1k} t_1+\omega_{2k} t_2+\phi_k\right)}\\
&=\sum_{k=1}^K \left[a_k {\rm cos}\left(\omega_{1k} t_1+\omega_{2k} t_2\right)-b_k {\rm sin}\left(\omega_{1k} t_1+\omega_{2k} t_2\right)\right]\\
&+j\; \left[a_k {\rm sin}\left(\omega_{1k} t_1+\omega_{2k} t_2\right)+b_k {\rm cos}\left(\omega_{1k} t_1+\omega_{2k} t_2\right)\right],\\
\end{split}
\end{equation}
where ${\bm{\theta}}_{2D}=\left[a_1,b_1,\omega_{11},\omega_{21},\cdots,a_K,b_K,\omega_{1K},\omega_{2K} \right]^T \in {\mathbb{R}}^{4K}$ is the unknown signal parameter vector, $a_k=A_k{\rm cos}\phi_k$, $b_k=A_k{\rm sin}\phi_k$, with $A_k$ and $\phi_k$ being the amplitude and phase of the $k$th sinusoidal component, and $t_l$ ($l=1,2$) is the time index for the $l$th dimension.

Similar to (\ref{signc}), the 2-D complex-valued signed measurements ${\bf Y}_{\rm 2D}=\{y_{n_1,n_2}\}\in \{1+j,1-j,-1+j,-1-j\}^{N_1\times N_2}$ can be obtained by comparing the noisy signal to the threshold ${\bf H}_{\rm 2D}=\{h_{n_1,n_2}\}\in \mathbb{C}^{N_1\times N_2}$:
\begin{equation}
y_{n_1,n_2}={\rm signc} \left(s_{n_1,n_2}\left(\bm{\theta}_{\rm 2D}\right)+e_{n_1,n_2}-h_{n_1,n_2}\right),
\end{equation}
where ${\bf E}_{\rm 2D}=\{e_{n_1,n_2}\}\in \mathbb{C}^{N_1\times N_2}$ is the unknown additive noise matrix with $e_{n_1,n_2}$ being i.i.d. circularly symmetric complex-valued Gaussian with zero-mean and unknown variance $\sigma^2$.

In the 2D complex-valued case, the $\left(i+1\right)$th MM iteration becomes
\begin{equation}
\begin{split}
\min_{\bm{\widetilde{\theta}}_{\rm 2D},\lambda} &g_{\rm 2D}\left({\bm{\widetilde{\theta}}_{\rm 2D},\lambda}|\bm{\widetilde{\beta}}_{\rm 2D}^i\right)=\min_{\bm{\widetilde{\theta}}_{\rm 2D},\lambda} \sum_{n_1=0}^{N_1-1}\sum_{n_2=0}^{N_2-1}\\
&\left|{ s}_{n_1,n_2}\left(\bm{\widetilde {\theta}}_{\rm 2D}\right)-\lambda h_{n_1,n_2}-{\widetilde z}_{n_1,n_2}\left(\bm{\widetilde {\beta}}_{\rm 2D}^i\right)\right|^2,\\
\label{UNKNOWN3}
\end{split}
\end{equation}
where $\bm{\widetilde {\beta}}_{\rm 2D}=\left[{\bm{\widetilde {\theta}}}_{\rm 2D}^T,\lambda\right]^T$ is the unknown parameter vector with $\lambda=\frac{1}{\sigma/\sqrt{2}}$, and ${\widetilde z}_{n_1,n_2}$ is defined similarly to (\ref{zupdate}). The cyclic algorithm for solving (\ref{UNKNOWN3}) can be obtained by replacing ${\bf h}$, ${\bf s}$ and ${\bf y}$ in (\ref{LAMBDA2}) by ${\rm vec}{\left({\bf H}_{\rm 2D}\right)}$, ${\rm vec}{\left({\bf S}_{\rm 2D}\right)}$ and ${\rm vec}{\left({\bf Y}_{\rm 2D}\right)}$, respectively, and by appropriately modifying the steps used to decrease $g_{\rm 2D}\left({\bm{\widetilde{\theta}}_{\rm 2D},\lambda_{m+1}^{i+1}}|\bm{\widetilde{\beta}}_{\rm 2D}^i\right)$ by means of 2-D FFT and 2-D CZT spectral zooms. We once again omit the details because the modifications are straightforward.

Finally, note that for 2-D complex-valued sinusoids, the 1bBIC cost function can be shown to be:
\begin{equation}
\begin{split}
&{\rm 1bBIC}\left({\breve K}\right)=\\
&-\!\!2\!\!\sum_{n_1=0}^{N_1-1}\!\!\sum_{n_2=0}^{N_2-1}\!\!\!{\rm log}\!\!\left(\!\!\Phi \!\! \left(\!\!{\rm Re}\!\left(y_{n_1,n_2}\right)\! \frac{{\rm Re}\!\left(\!\!s_{n_1,n_2}\left(\bm{\widehat{\theta}}\right)\!\right)\!\!-\!\!{\rm Re}\!\left(h_{n_1,n_2}\right)}{\widehat{\sigma}/\sqrt{2}}\!\right)\!\!\right)\\
&-\!\!2\!\!\sum_{n_1=0}^{N_1-1}\!\!\sum_{n_2=0}^{N_2-1}\!\!\!{\rm log}\!\!\left(\!\!\Phi \!\! \left(\!\!{\rm Im}\!\left(y_{n_1,n_2}\right)\! \frac{{\rm Im }\!\left(\!\!s_{n_1,n_2}\left(\bm{\widehat{\theta}}\right)\!\right)\!\!-\!\!{\rm Im}\!\left(h_{n_1,n_2}\right)}{\widehat{\sigma}/\sqrt{2}}\!\right)\!\!\right)\\
&+\!\!6\breve{K}{\rm log}N_1N_2.\\
\end{split}
\label{BIC2D}
\end{equation}
\section{Simulated and Experimental Examples}
In this section, we present both simulation and experimental examples to demonstrate the performance of the proposed algorithms for estimating the sinusoidal parameters and for determining the number of sinusoids using signed measurements. We start by presenting numerical examples of 1-D sinusoidal parameter estimation from real-valued signed measurements obtained via one-bit sampling with both time-varying and fixed non-zero thresholds. The proposed 1bMMRELAX algorithm is compared with 1bCLEAN and 1bRELAX in terms of estimation accuracy and computational complexity, and the model order determination performance of 1bBIC is also evaluated. Then, we present experimental examples of using 1bMMRELAX with 1bBIC for range-Doppler imaging using measured automotive radar data. All the examples were run on a PC with 3.10 GHz CPU and 16.00 GB RAM.

\subsection{Implementation Details}
In our implementation of 1bMMRELAX, we terminate the MM iterations if the relative change of the negative log-likelihood function $l\left(\widetilde{\bm{\beta}}\right)$ between two consecutive iterations is less than $10^{-5}$ or a maximum number of the MM iterations $I_M=30$ is reached. Within each MM iteration, we terminate the inner loop if the relative change of the objective function in (\ref{UNKNOWN}) is less than $10^{-5}$. $N_1$ is set as the smallest power of 2 larger than or equal to $N$, and $N_2$ is set such that the length of FFTs and IFFTs included in CZTs is $2N_1$. Additionally, we terminate the update iterations of 1bRELAX for each model order when the relative change of $l\left(\widetilde{\bm{\beta}}\right)$ is less than $10^{-5}$ or the maximum number of the update iterations $I_R=30$ is reached.

\begin{figure}
\centering
\subfigure[]{
\label{(MSE1)}
\includegraphics[width=3in,height=1.5in]{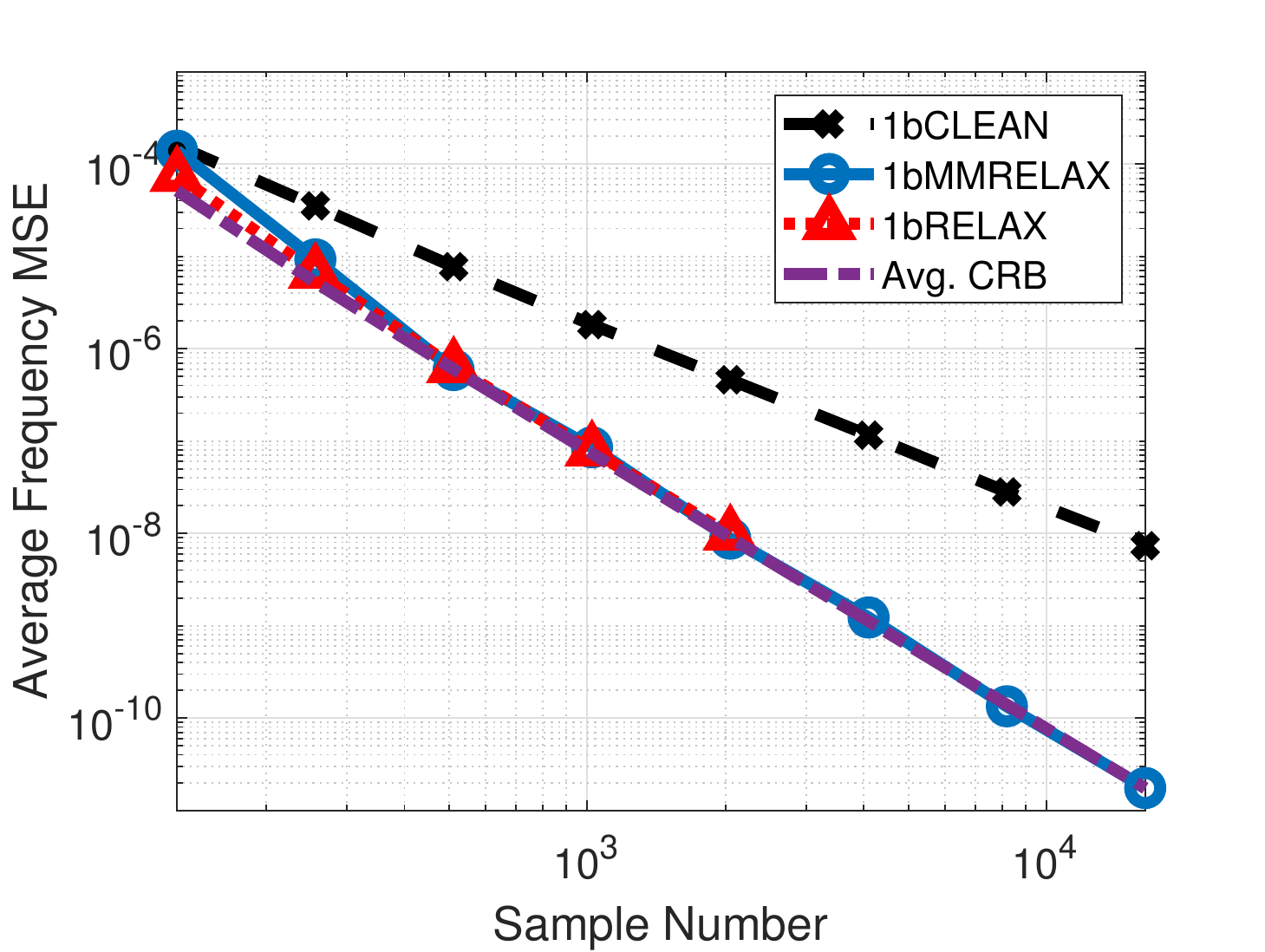}}
\subfigure[]{
\label{(MSE3)}
\includegraphics[width=3in,height=1.5in]{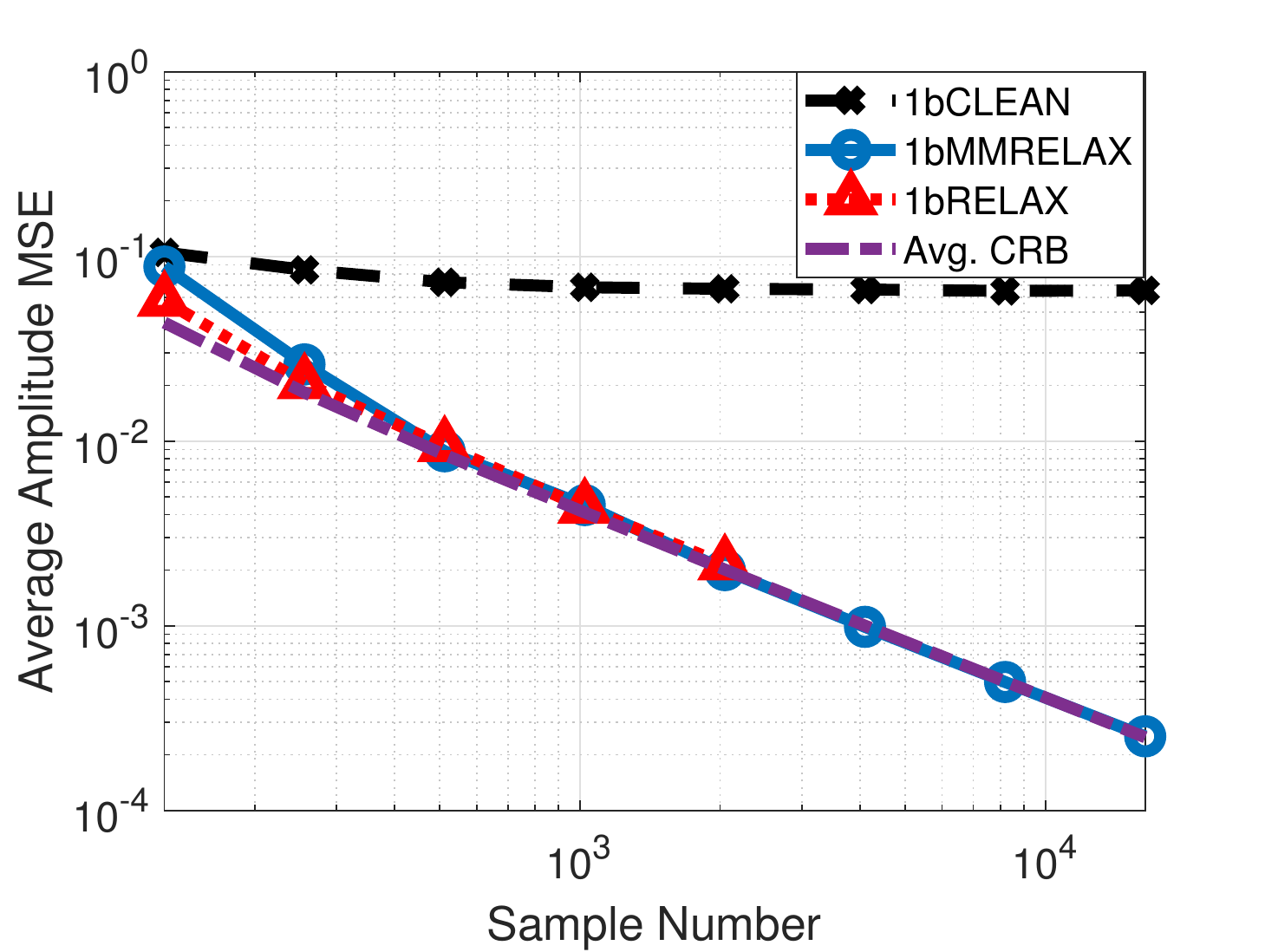}}
\subfigure[]{
\label{(MSE2)}
\includegraphics[width=3in,height=1.5in]{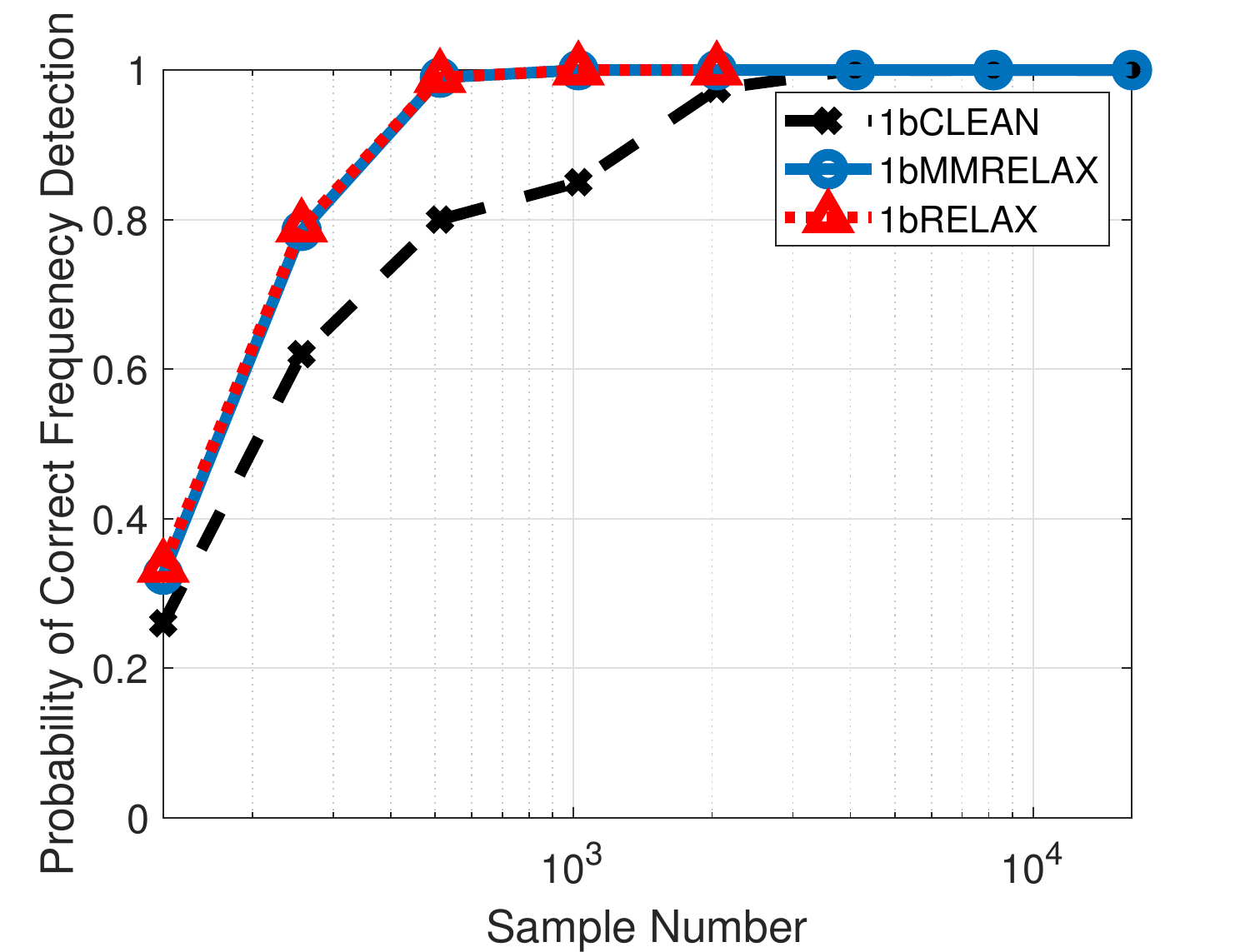}}
\caption{Sinusoidal parameter estimation performance of 1bCLEAN, 1bMMRELAX, and 1bRELAX for a time-varying threshold and varying sample lengths $N$ when SNR = 10 dB: (a) average frequency MSEs vs. $N$, (b) average amplitude MSEs vs. $N$, and (c) probabilities of correct frequency detection vs. $N$.}
\label{Exam11}
\end{figure}
\begin{figure}
\centering
\subfigure[]{
\label{(MSE21)}
\includegraphics[width=3in,height=1.5in]{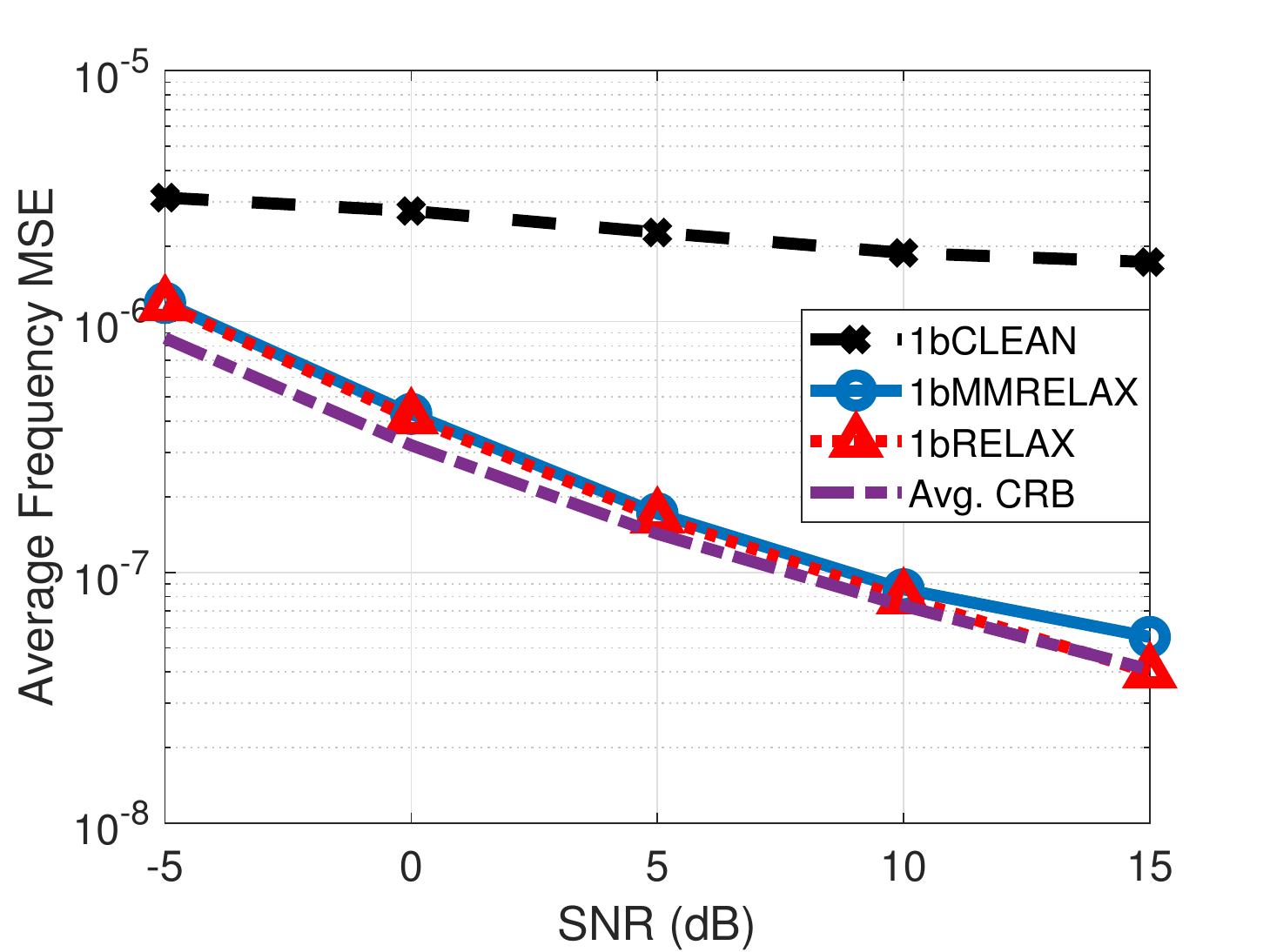}}
\subfigure[]{
\label{(MSE23)}
\includegraphics[width=3in,height=1.5in]{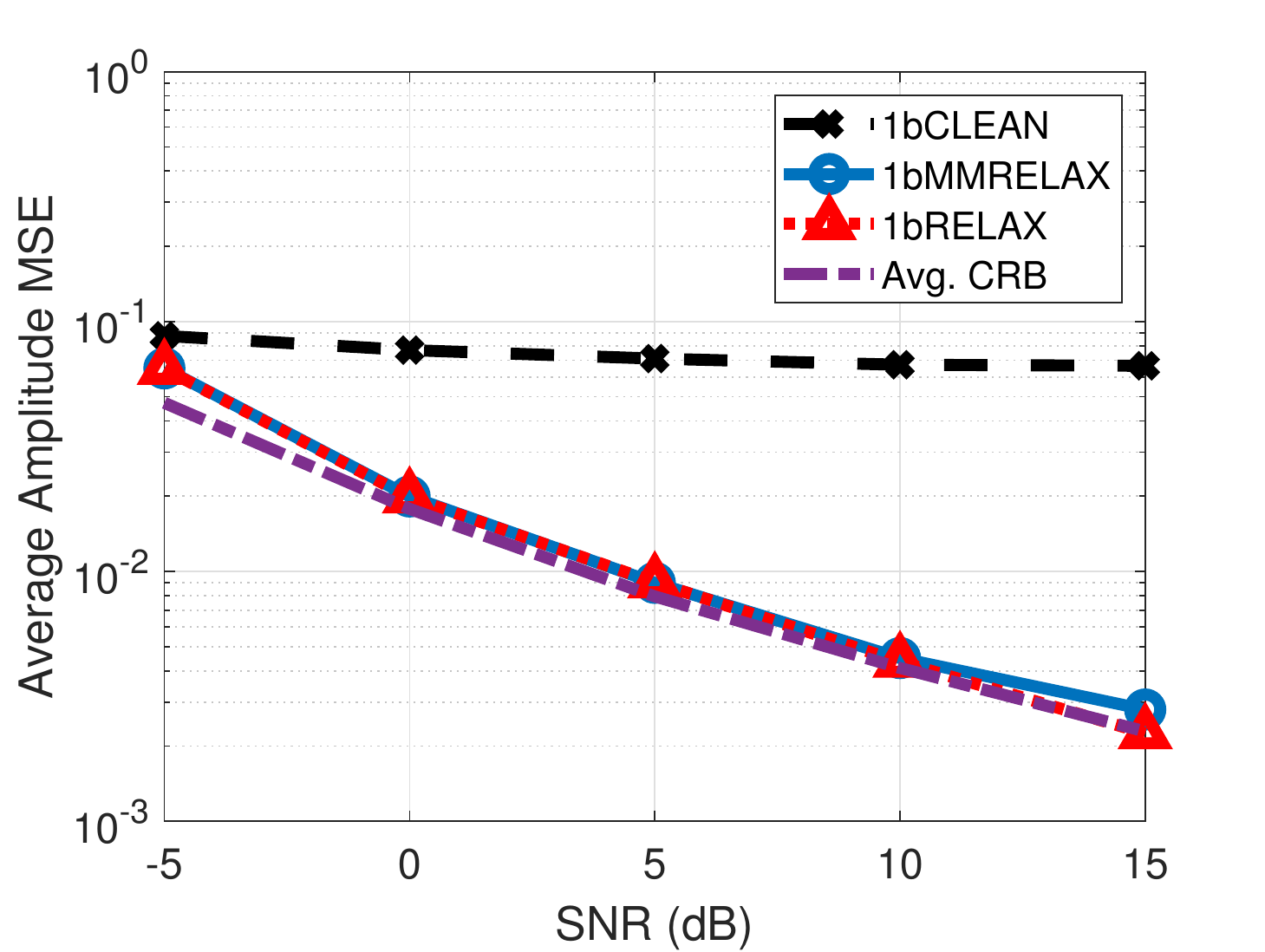}}
\subfigure[]{
\label{(MSE22)}
\includegraphics[width=3in,height=1.5in]{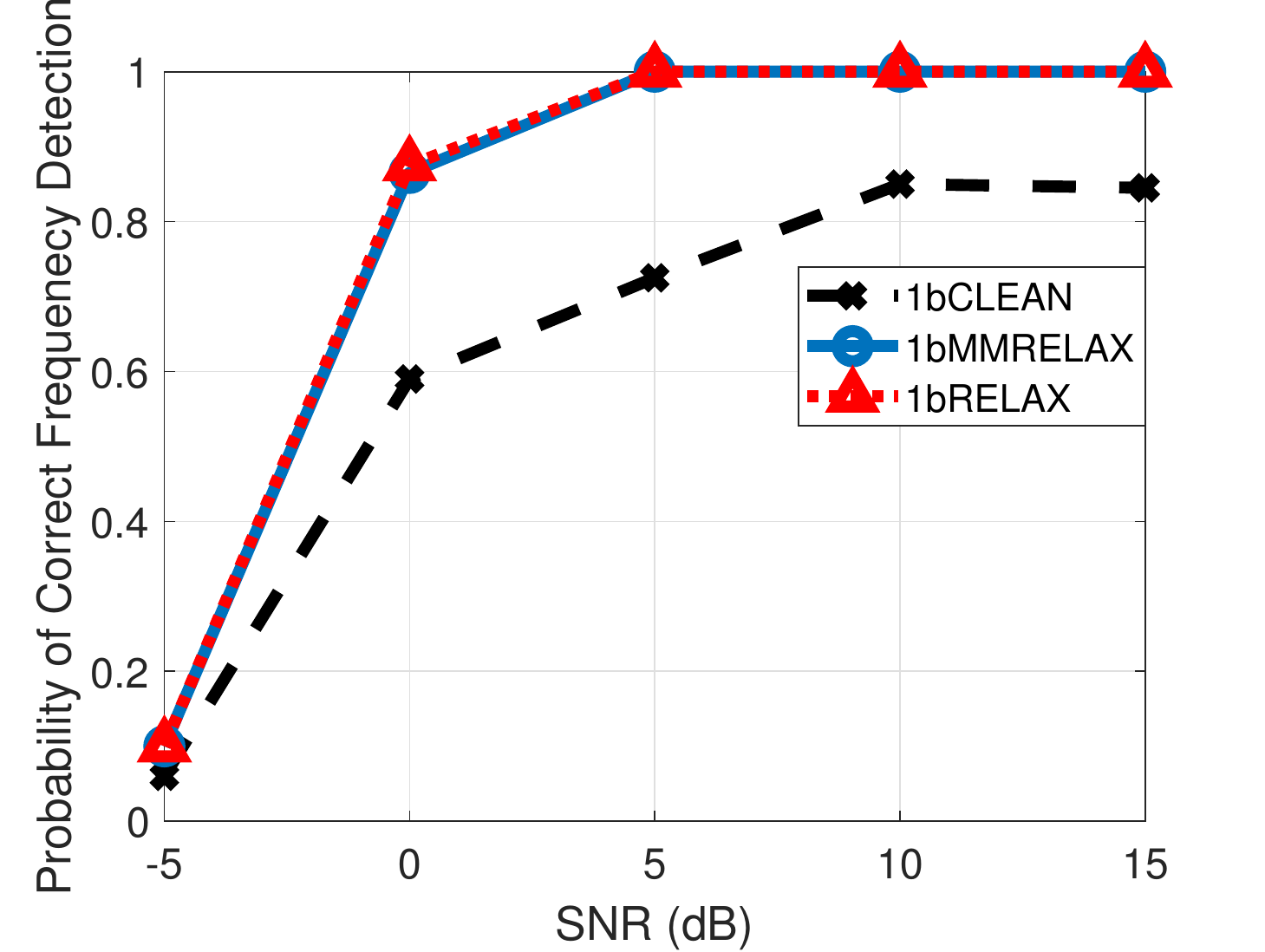}}
\caption{Sinusoidal parameter estimation performance of 1bCLEAN, 1bMMRELAX, and 1bRELAX for a time-varying threshold and varying SNRs when $N=1024$: (a) average frequency MSEs vs. SNR, (b) average amplitude MSEs vs. SNR, and (c) probabilities of correct frequency detection vs. SNR.}
\label{Exam12}
\end{figure}

\begin{figure}
\centering
\includegraphics[width=3in,height=1.5in]{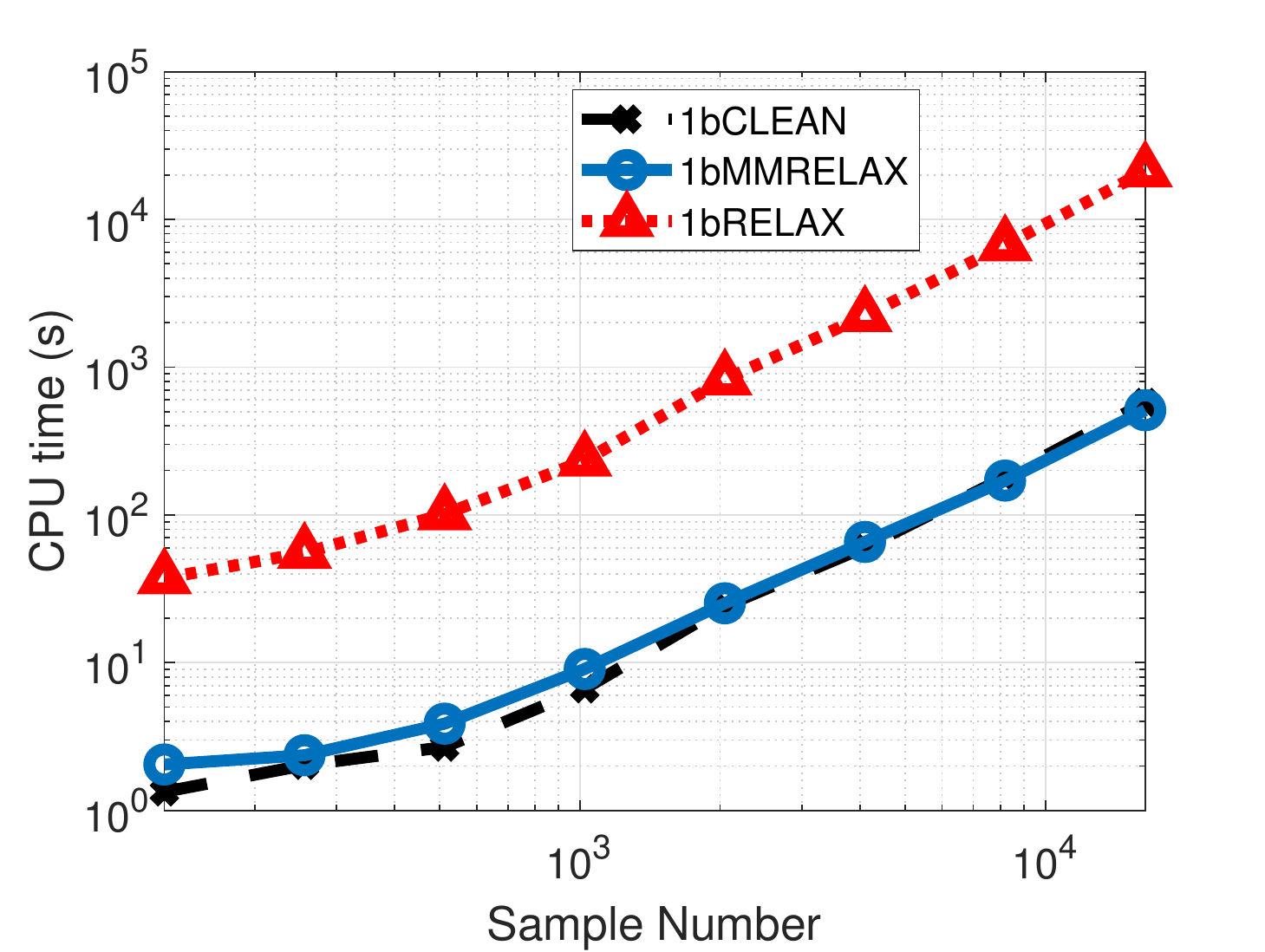}
\caption{Average running time vs. $N$ for 1bCLEAN, 1bRELAX, and 1bMMRELAX for SNR=10 dB.}
\label{Time}
\end{figure}




\subsection{Sinusoidal Parameter Estimation}
\subsubsection{Time-Varying Threshold}
We first study the estimation accuracy and the computational complexity of the 1bMMRELAX algorithm when compared to 1bCLEAN and 1bRELAX.

\textit{Example 1}:
We consider a signal composed of $K=6$ sinusoids with frequencies $\left\{0.11\times 2\pi,\! \right.$ $\left.\left(0.11\!+\!1/N\right)\times 2\pi,\! 0.2\times2\pi,\! 0.3\times 2\pi,\! 0.37\times 2\pi,\! 0.45\times 2\pi\right\}$, amplitudes $\left\{1, 1, 0.7, 0.8, 0.6, 0.5\right\}$, and phases $\left\{7\pi/6, \pi/6,\right.$ $ \left. \pi/2,\pi/4, 11\pi/6, \pi\right\}$. Note that the first two frequencies are closely spaced. A time-varying threshold, which is randomly generated from a discrete set of 8 values uniformly distributed over $\left[-1,1\right]$, is used to obtain the signed measurements. To evaluate the estimation accuracy, the average mean-squared errors (MSEs) of the sinusoidal parameter estimates are estimated from 200 Monte Carlo runs. Note that each trial corresponds to an independent noise and threshold realization.

We plot the average frequency and amplitude MSEs, the corresponding CRBs, as well as the probabilities of correct frequency detection $P_d$ as functions of $N$ and SNR in Fig. \ref{Exam11} and Fig. \ref{Exam12}, respectively. In each trial, the frequency detection is considered correct if the maximum absolute error of the frequency estimates is less than $\frac{2\pi}{N}$. Note that the missed detection trials are not taken into account in the computation of the average MSEs. Note also that since it is too slow to perform 200 Monte Carlo trails for 1bRELAX for large $N$ scenarios, 1bRELAX for $N>2048$ is not shown in Fig. \ref{Exam11}, where SNR = 10 dB. Inspecting these results for various $N$ in Fig. \ref{Exam11}, we see that 1bRELAX and 1bMMRELAX always provide lower MSEs and higher $P_d$ values than 1bCLEAN. Specifically, 1bCLEAN cannot provide a good amplitude estimation performance (especially for the first two closely spaced sinusoids) even when $N$ is large. The MSEs of the estimates obtained by using 1bRELAX and 1bMMRELAX are close to the CRBs when $N\ge 512$. Furthermore, the results in Fig. \ref{Exam12}, where $N=1024$, demonstrate that 1bMMRELAX and 1bRELAX have better noise tolerance than 1bCLEAN and provide excellent estimation performance when SNR$>0$ dB. Additionally, as $N$ increases, $P_d$ goes to 1, and the increase of SNR results in an improvement in $P_d$ for all three algorithms, as expected.

The average computation times needed by the aforementioned algorithms, obtained using 5 Monte Carlo trials are recorded in seconds and plotted on a logarithmic scale ($10\log_{10}(\cdot)$) in Fig. \ref{Time} for various $N$ when the SNR is 10 dB. The required times for other SNR values are similar to those for SNR = 10 dB. As predicted, 1bMMRELAX and 1bCLEAN require similar computational times, and both are much faster than 1bRELAX. Specifically, 1bMMRELAX is more than an order of magnitude faster than 1bRELAX, while maintaining similar sinusoidal parameter estimation accuracy. Regarding certain penalized/sparse algorithms proposed in the literature for sinusoidal parameter estimation from signed measurements \cite{CLJP162,CLJP17}, we do not consider them in this comparative study because they are slow for large values of $N$ and have worse performance than 1bRELAX \cite{CLJP17}.

\begin{figure}
\centering
\subfigure[]{
\label{(EXAM61)}
\includegraphics[width=3in,height=1.5in]{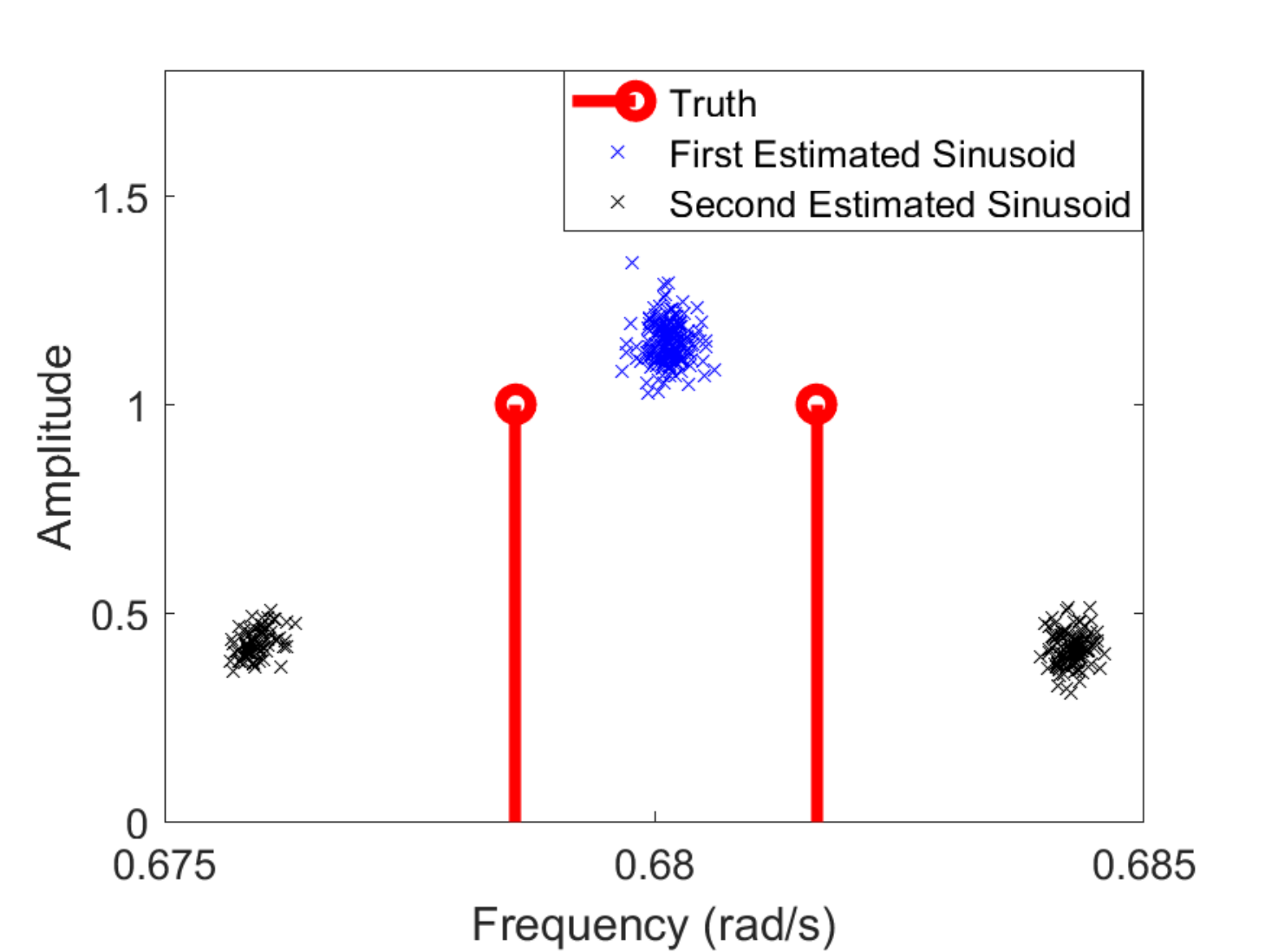}}
\subfigure[]{
\label{(EXMA62)}
\includegraphics[width=3in,height=1.5in]{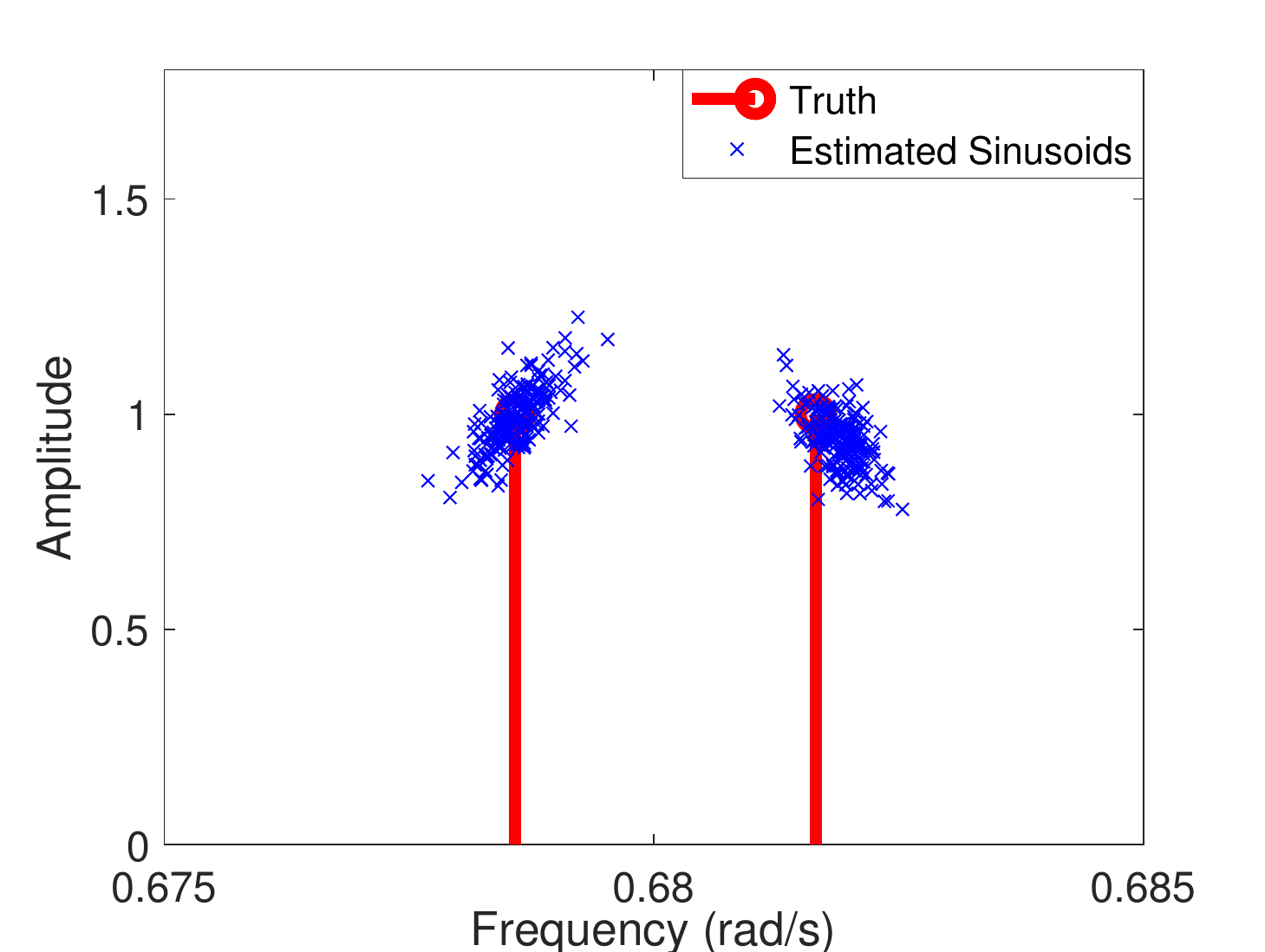}}
\subfigure[]{
\label{(EXAM63)}
\includegraphics[width=3in,height=1.5in]{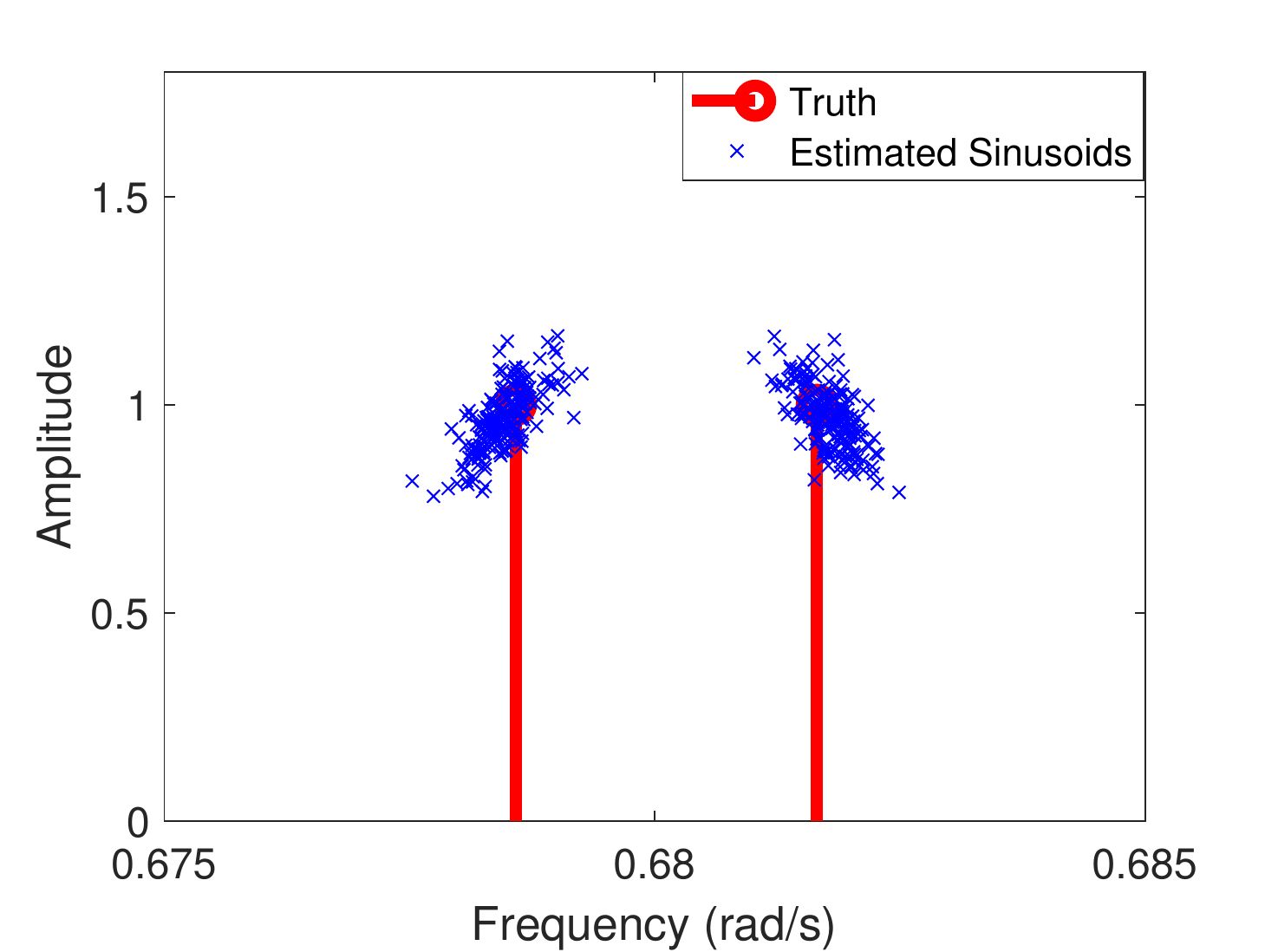}}
\caption{Frequency spectra of (a) 1bCLEAN, (b) 1bMMRELAX, (c) 1bRELAX for two closely spaced sinusoids in 200 Monte Carlo trials, for $N=1024$ and SNR = 10 dB, and a time-varying threshold.}
\label{Exam2}
\end{figure}

\textit{Example 2}:
To further illustrate the resolution capability and estimation accuracy of 1bCLEAN, 1bMMRELAX and 1bRELAX, we consider a signal composed of two sinusoids with a small frequency separation and the same amplitudes and phases. The parameters of the two sinusoids are: $\omega_1=0.108\times 2\pi$, $\omega_2=\left(0.108+\frac{1}{2N}\right)\times 2\pi$, $A_1=A_2=1$, and $\phi_1=\phi_2=\frac{1}{3}\pi$. We set $N=1024$ and SNR = 10 dB. The same type of time-varying threshold as in Example 1 is used here to obtain the signed measurements. The two sinusoids are separated by only $\frac{\pi}{N}$ and therefore it is a challenging task for 1bCLEAN to resolve them. We show the simulation results of 200 Monte Carlo runs in Fig. \ref{Exam2}. Specifically, in Fig. \ref{Exam2}(a), the blue scatterers and black scatterers show the sinusoids obtained via the first step and second step of 1bCLEAN, respectively. Inspecting the results, we can see that 1bCLEAN cannot properly resolve the two sinusoids, whereas 1bRELAX and 1bMMRELAX can resolve them and thus possess super resolution capability.

In sum, taking into account the resolution capability, estimation accuracy, noise tolerance, and computational complexity, the examples suggest that 1bMMRELAX is preferred over 1bCLEAN and 1bRELAX. We also remark that as the number of sinusoids increases, the computational complexity reduction offered by 1bMMRELAX over 1bRELAX becomes even more significant.

\subsubsection{Fixed Non-zero Threshold}
\begin{figure}
\centering
\subfigure[]{
\label{(MSE31)}
\includegraphics[width=3in,height=1.5in]{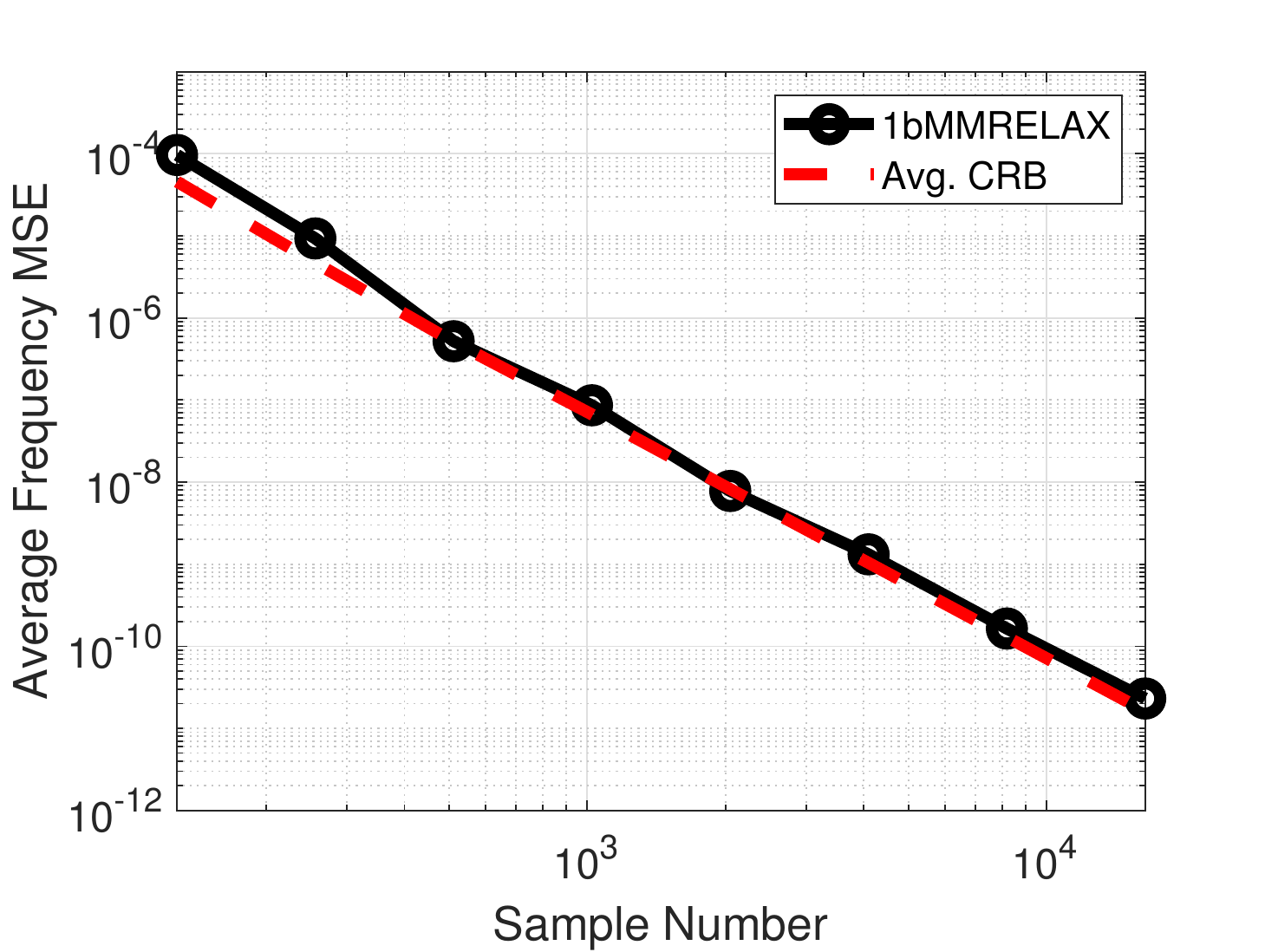}}
\subfigure[]{
\label{(MSE33)}
\includegraphics[width=3in,height=1.5in]{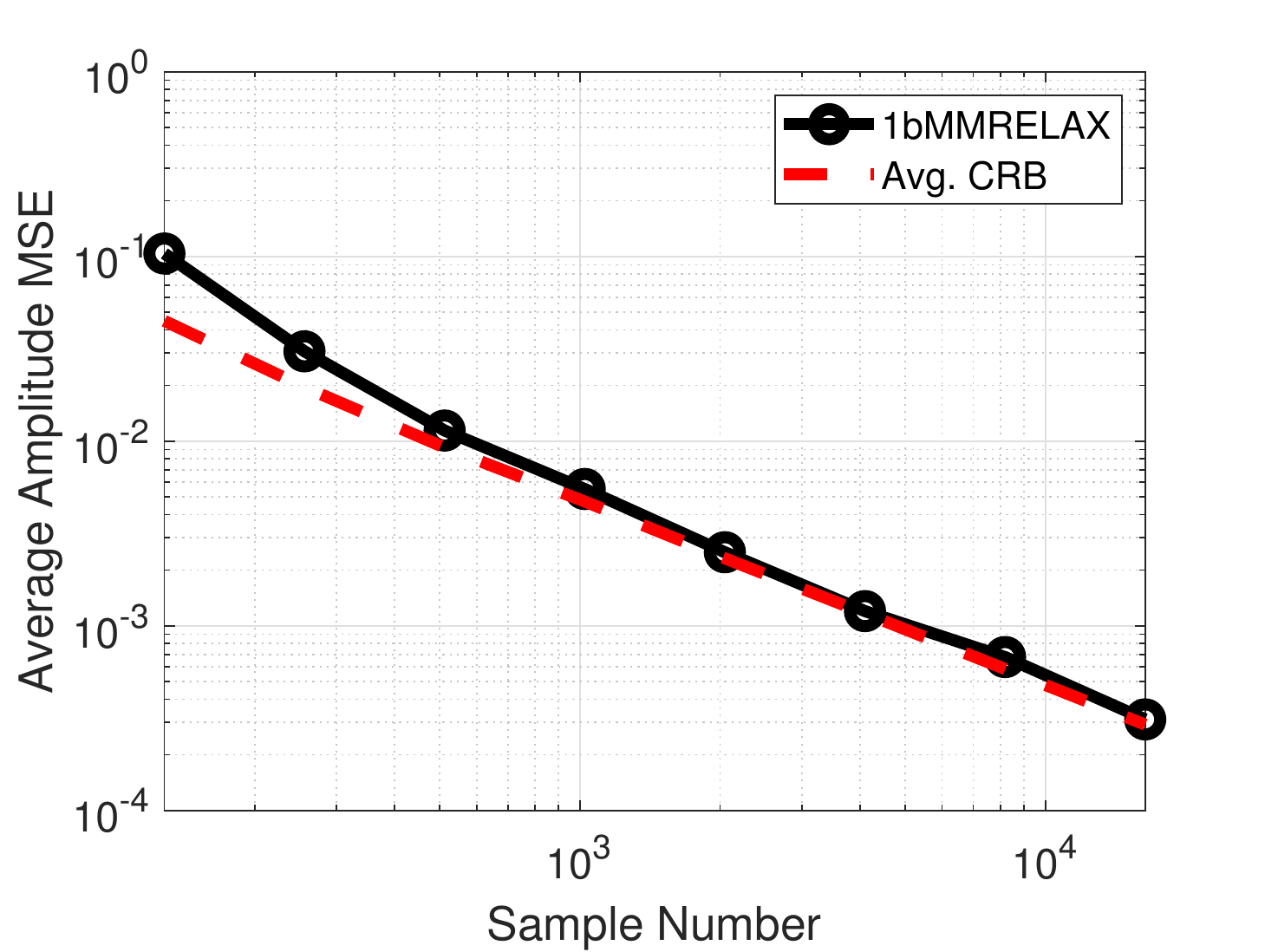}}
\subfigure[]{
\label{(MSE32)}
\includegraphics[width=3in,height=1.5in]{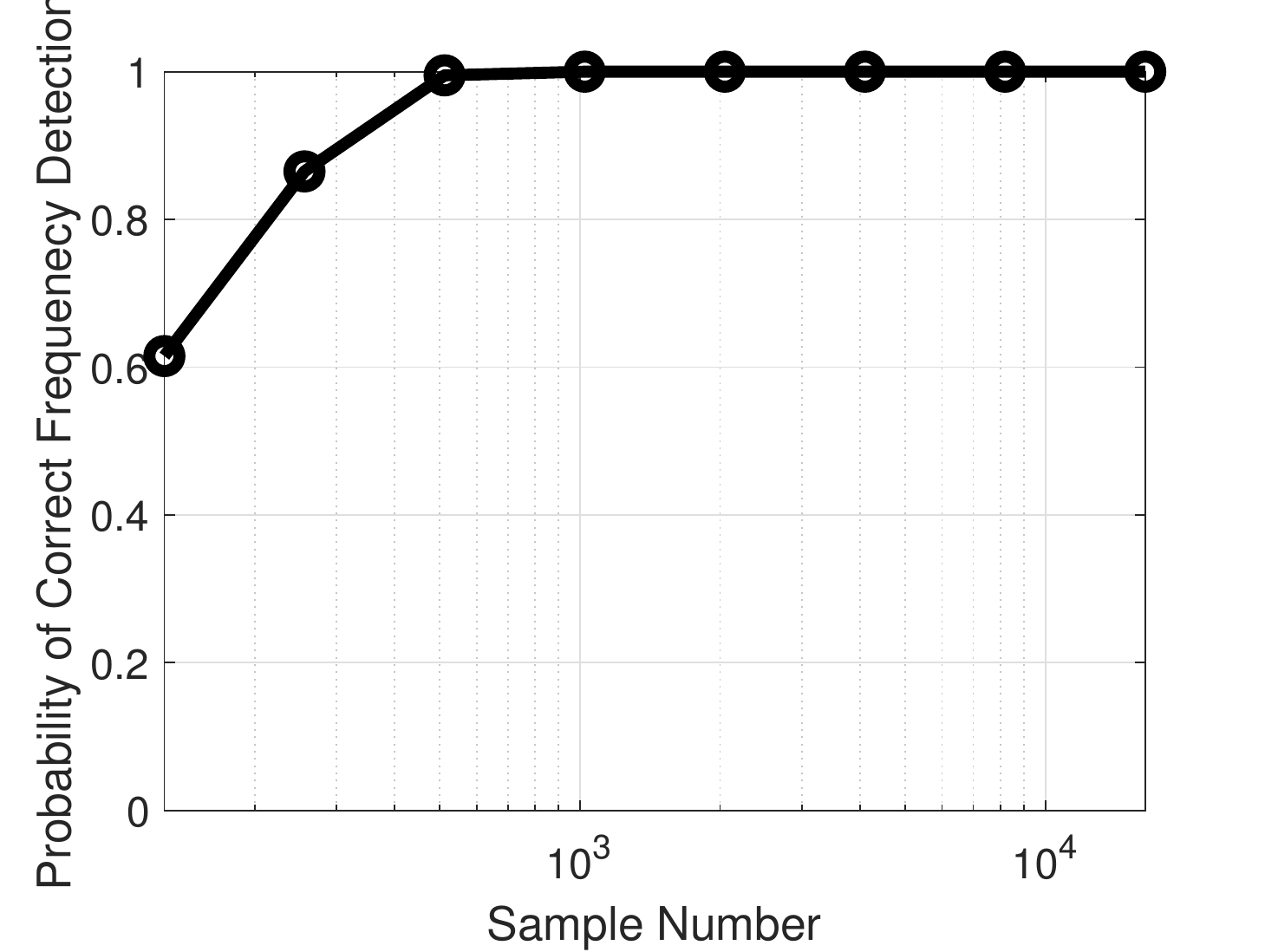}}
\caption{Sinusoidal parameter estimation performance of 1bMMRELAX for a fixed non-zero threshold and varying sample lengths $N$ when SNR = 10 dB: (a) average frequency MSEs vs. $N$, (b) average amplitude MSEs vs. $N$, and (c) probabilities of correct frequency detection vs. $N$.}
\label{FIX1}
\end{figure}

\begin{figure}
\centering
\subfigure[]{
\label{(MSE51)}
\includegraphics[width=3in,height=1.5in]{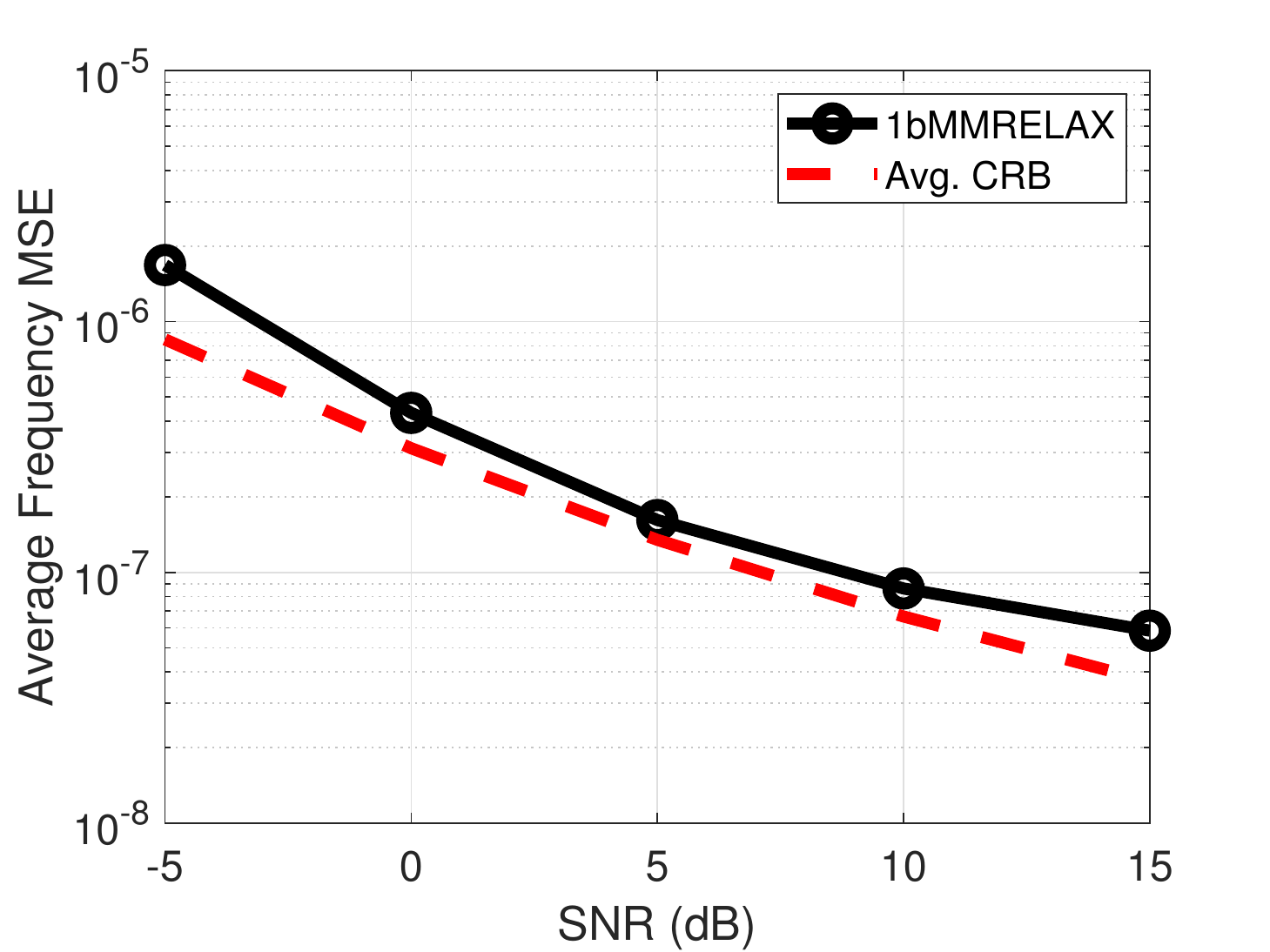}}
\subfigure[]{
\label{(MSE53)}
\includegraphics[width=3in,height=1.5in]{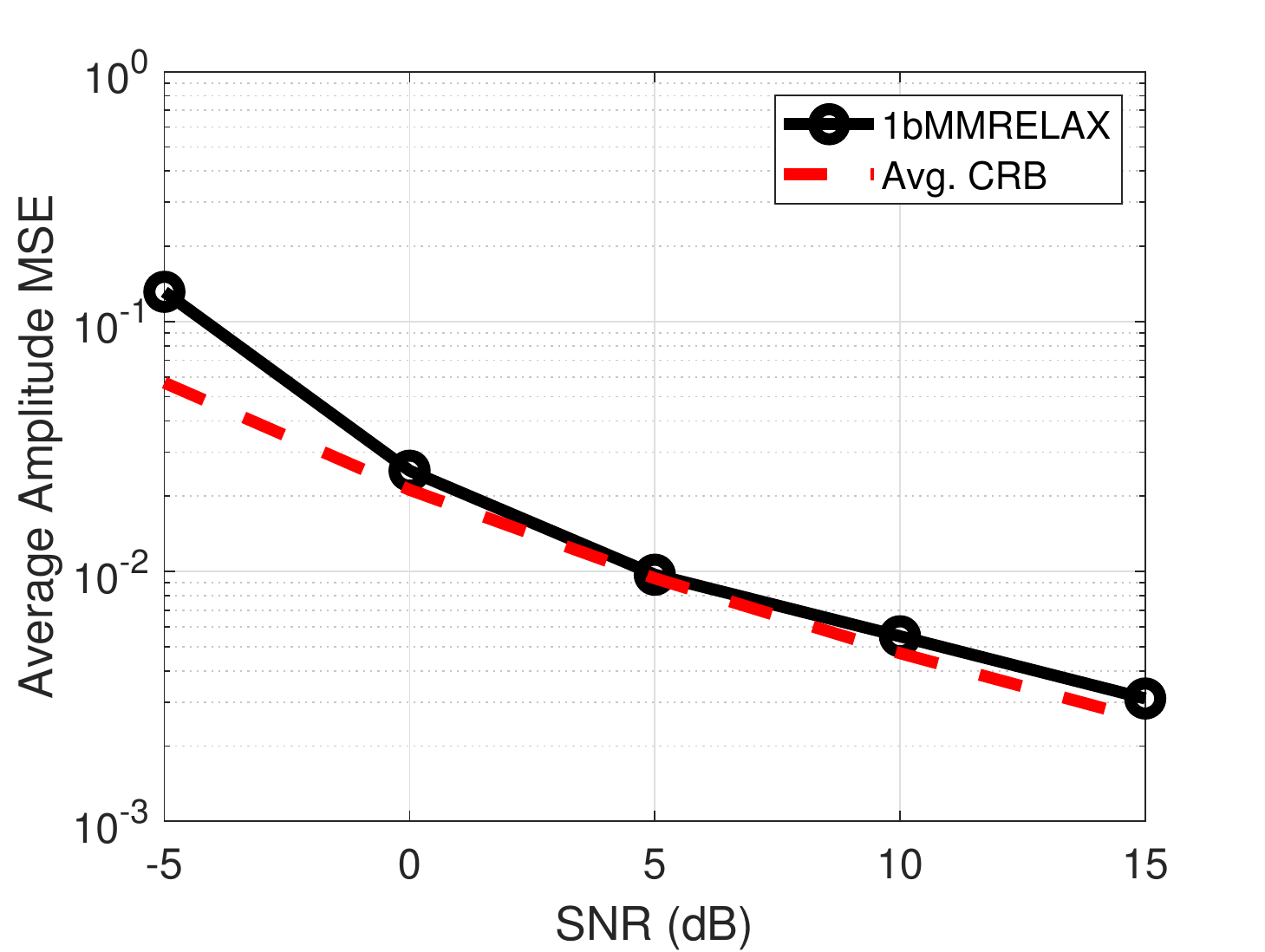}}
\subfigure[]{
\label{(MSE52)}
\includegraphics[width=3in,height=1.5in]{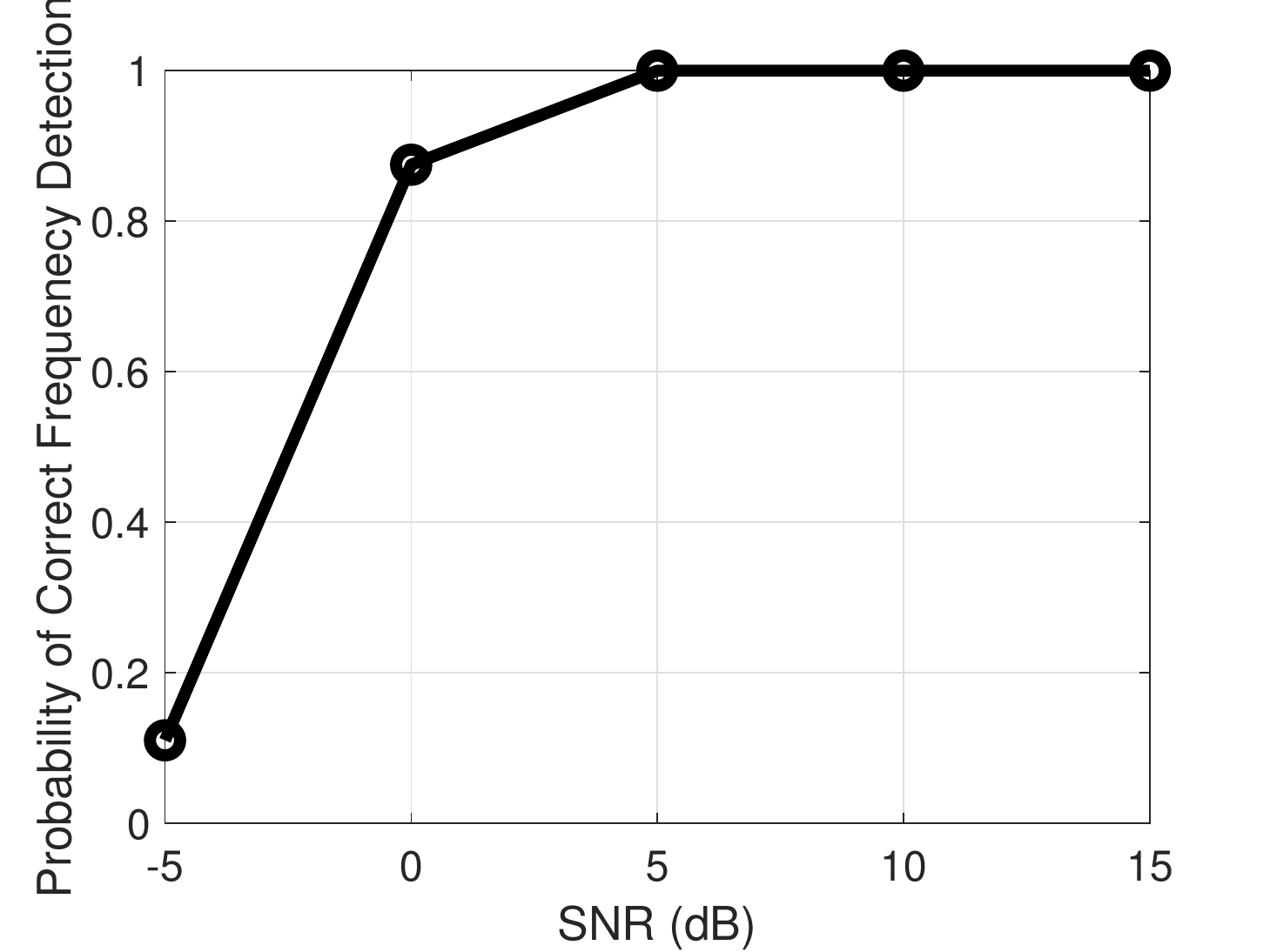}}
\caption{Sinusoidal parameter estimation performance of 1bMMRELAX for a fixed non-zero threshold and varying SNRs when $N=1024$: (a) average frequency MSEs vs. SNR, (b) average amplitude MSEs vs. SNR, and (c) probabilities of correct frequency detection vs. SNR.}
\label{FIX2}
\end{figure}

We next consider using the low cost fixed non-zero threshold to obtain the signed measurements, assuming that the signal does not have a DC component. The signal is the same as in Example 1. The fixed non-zero threshold is $\{h_n=0.5\}_{n=1}^N$. Fig. \ref{FIX1} and Fig. \ref{FIX2} show the average frequency and amplitude MSEs as well the probabilities of correct frequency detection $P_d$ over 200 Monte Carlo trials as a function of $N$ (when SNR = 10dB) and SNR (when N = 1024), respectively. As mentioned before, the MSEs are only for the correctly detected cases (see Example 1 for details). Inspecting these results, it can be seen that using the fixed non-zero threshold can yield a good frequency and amplitude estimation performance when $N\ge512$ and SNR$>0$ dB. Compared with using the time-varying threshold, using the fixed non-zero threshold saves hardware cost, while providing similar frequency and amplitude MSEs as well as similar $P_d$ values. The caveat is that the choice of the fixed threshold should be done more carefully than that of the time-varying threshold, as it might be expected.

\subsubsection{Model Order Determination via 1bBIC}
\begin{figure}
\centering
\includegraphics[width=3in,height=1.5in]{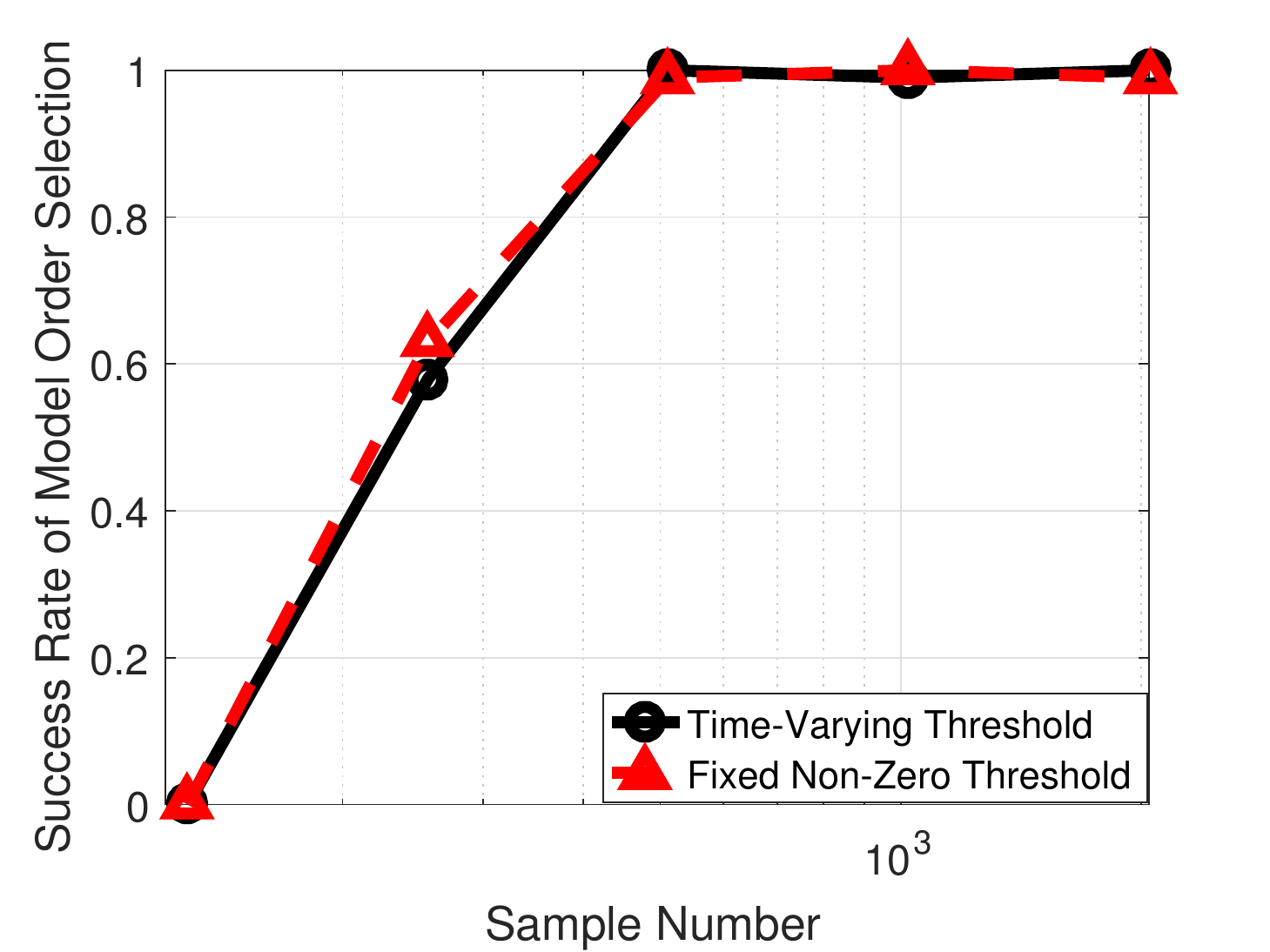}
\caption{Success rates of correct order determination vs. $N$ for time-varying as well as fixed non-zero thresholds when SNR = 10 dB.}
\label{Model1}
\end{figure}

\begin{figure}
\centering
\includegraphics[width=3in,height=1.5in]{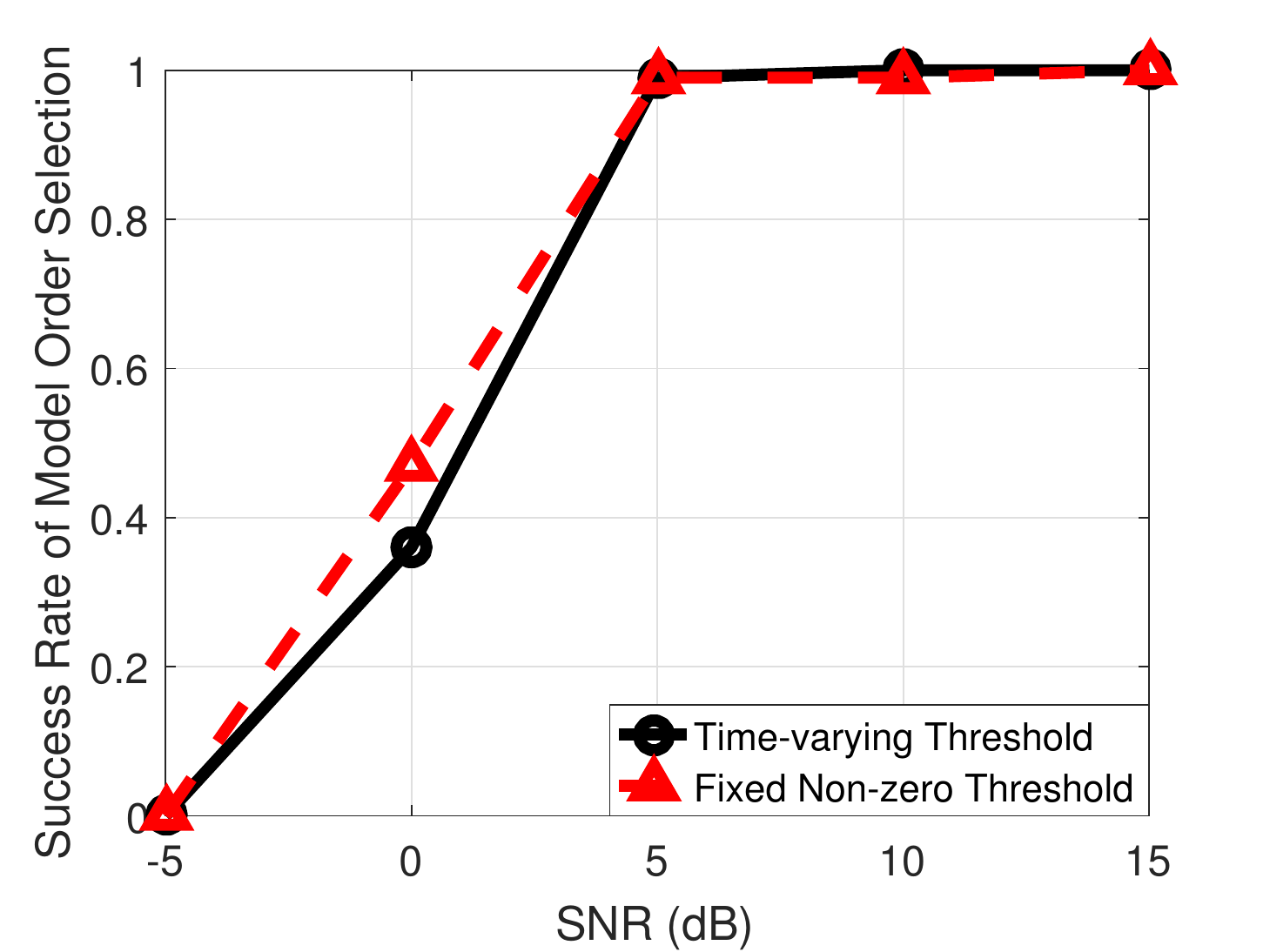}
\caption{Success rates of correct order determination vs. SNR for time-varying as well as fixed non-zero thresholds when $N=1024$.}
\label{Model2}
\end{figure}

We finally test the model order determination performance of 1bMMRELAX with 1bBIC. Signed measurements obtained with both fixed non-zero and time-varying thresholds are utilized in this experiment, and the same signal as in Example 1 is considered. Fig. \ref{Model1} and Fig. \ref{Model2}, respectively, show the success rates of correct order determination as a function of $N$ (when SNR = 10 dB) and SNR (when N = 1024). It can be seen that 1bMMRELAX with 1bBIC provides accurate order estimates as $N$ or SNR
increases. Again, similar performances are obtained using fixed non-zero threshold and time-varying threshold.

Note that all of the above experiments have considered the unknown $\sigma$ case. The performance improves slightly when $\sigma$ is known, as expected, but we will not show the results for known $\sigma$ in this paper as they are of a somewhat limited practical interest.

\subsection{Range-Doppler Imaging}

We use experimental data from a 24 GHz radar sensor to demonstrate the performance of 1bMMRELAX with 1bBIC for range-Doppler imaging from 2-D complex-valued signed measurements. The radar sensor, which is placed on a pedestrian bridge over the road, transmits periodic linear frequency modulated continuous waveform (LFMCW) sequences, with bandwidth $B=25 $ MHz and pulse repetition interval $T=80 $ $\mu$s. The received signal is sampled by high-precision ADCs. The measured data contain additive noise with unknown noise variance. The signed measurements are obtained by comparing the original high-precision data with a fixed non-zero threshold ${\bf H}=\{h_{n_1,n_2}=v_1+jv_2\}$, which can be implemented using a hardware similar to the common zero threshold case. The dimensions of the 2-D data matrix are $N_1=64$ and $N_2=512$. For illustration purposes, assuming that $A$ is the maximum of the modulus of the original signal, $v_1$ and $v_2$ are both chosen as $\frac{A}{4}$. Note that in practical applications where $A$ will be unknown, we can chose proper $v_1$ and $v_2$ based on the dynamic range of the radar receiver.

\begin{figure}
\centering
\subfigure[]{
\label{(EXAM81)}
\includegraphics[width=3in,height=1.5in]{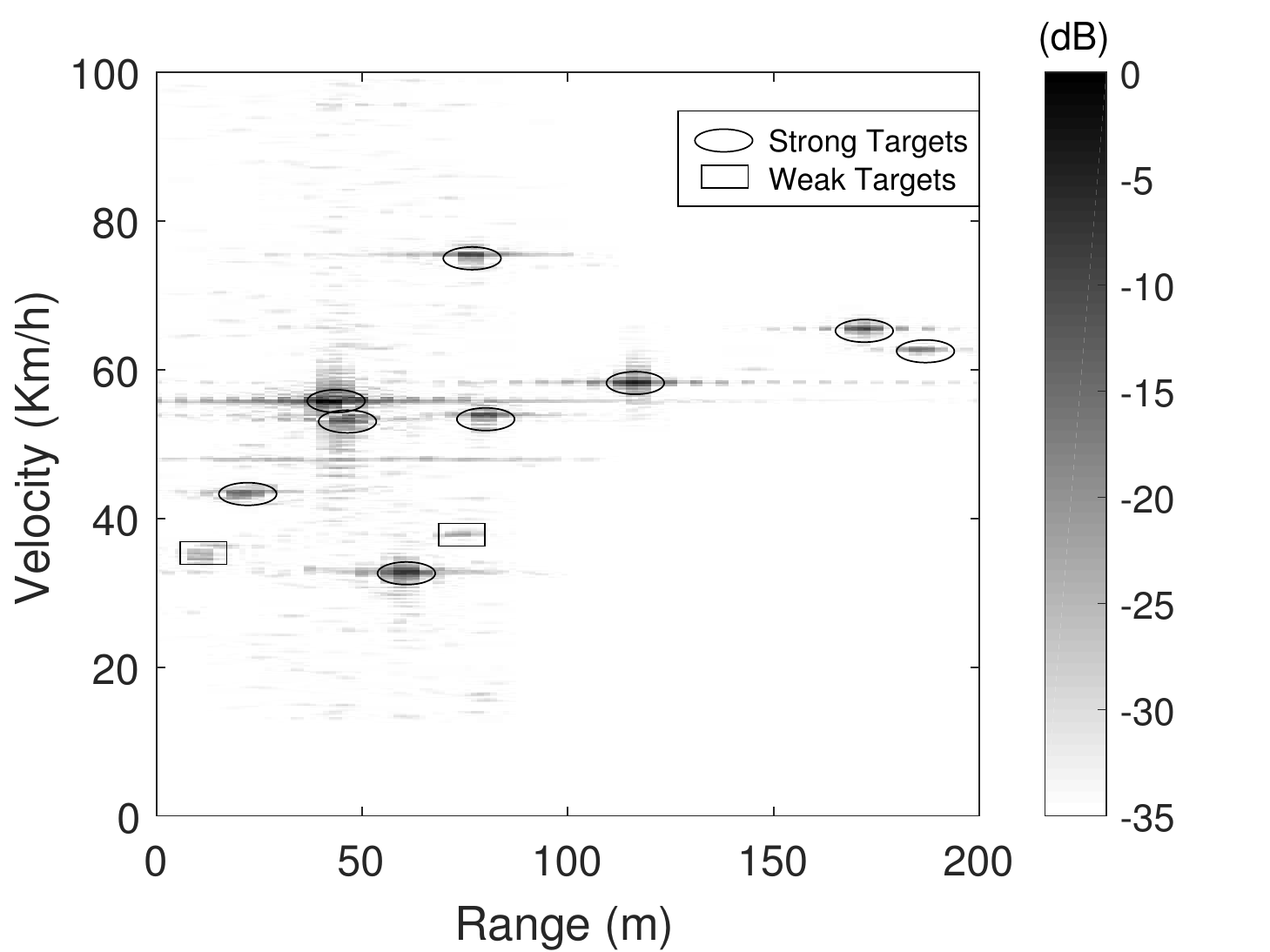}}
\subfigure[]{
\label{(EXMA82)}
\includegraphics[width=3in,height=1.5in]{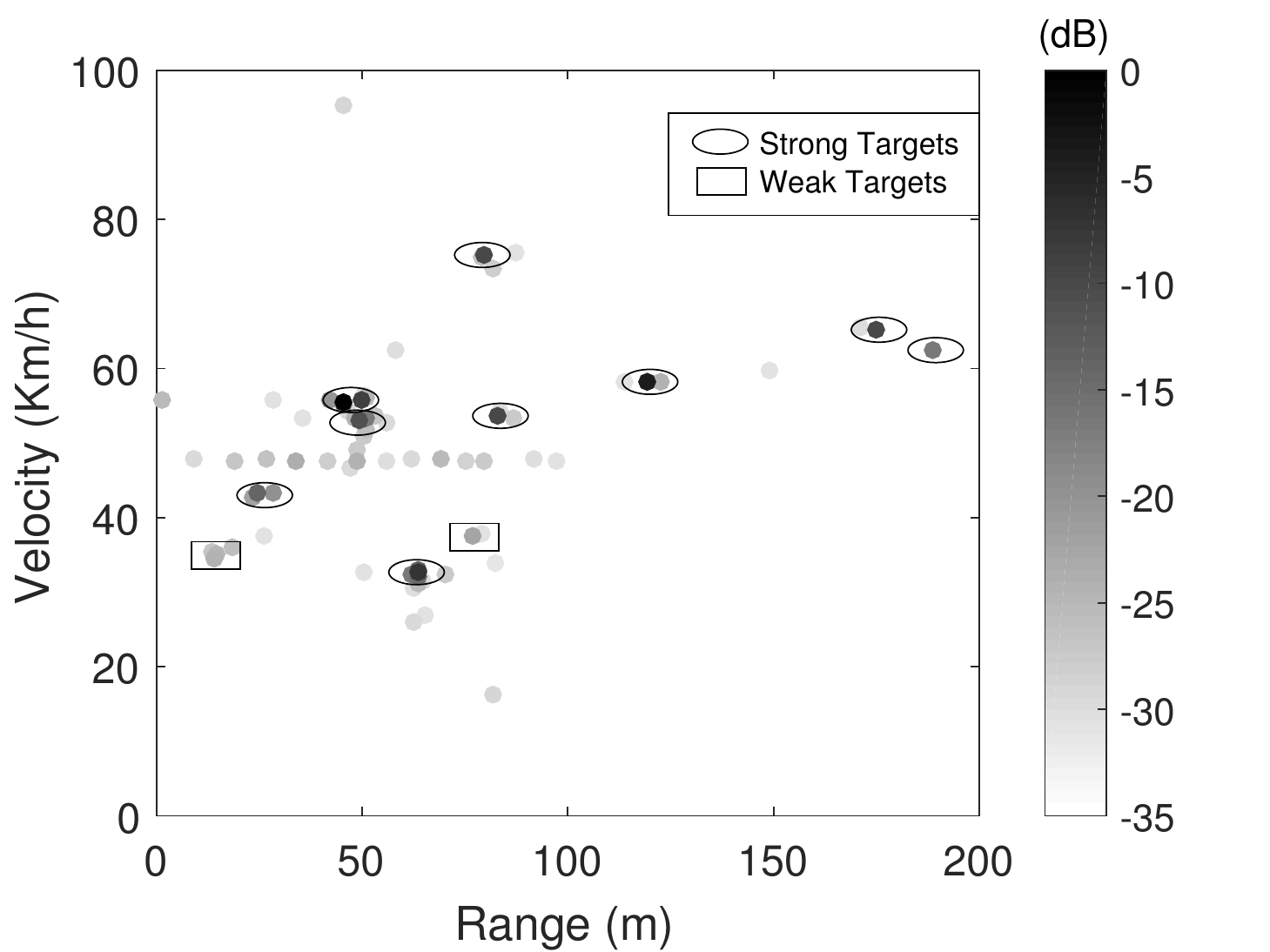}}
\subfigure[]{
\label{(EXMA83)}
\includegraphics[width=3in,height=1.5in]{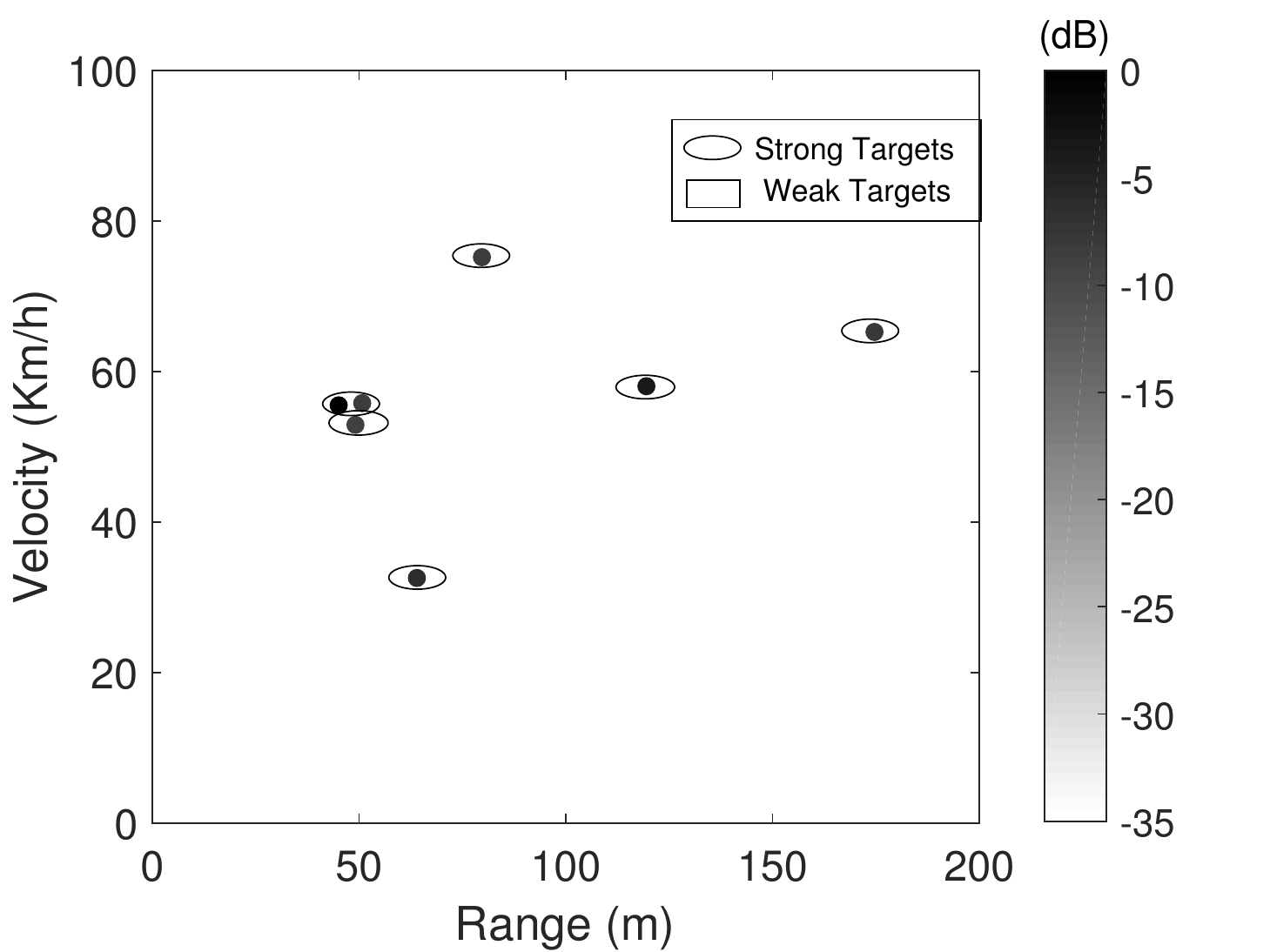}}
\subfigure[]{
\label{(EXAM84)}
\includegraphics[width=3in,height=1.5in]{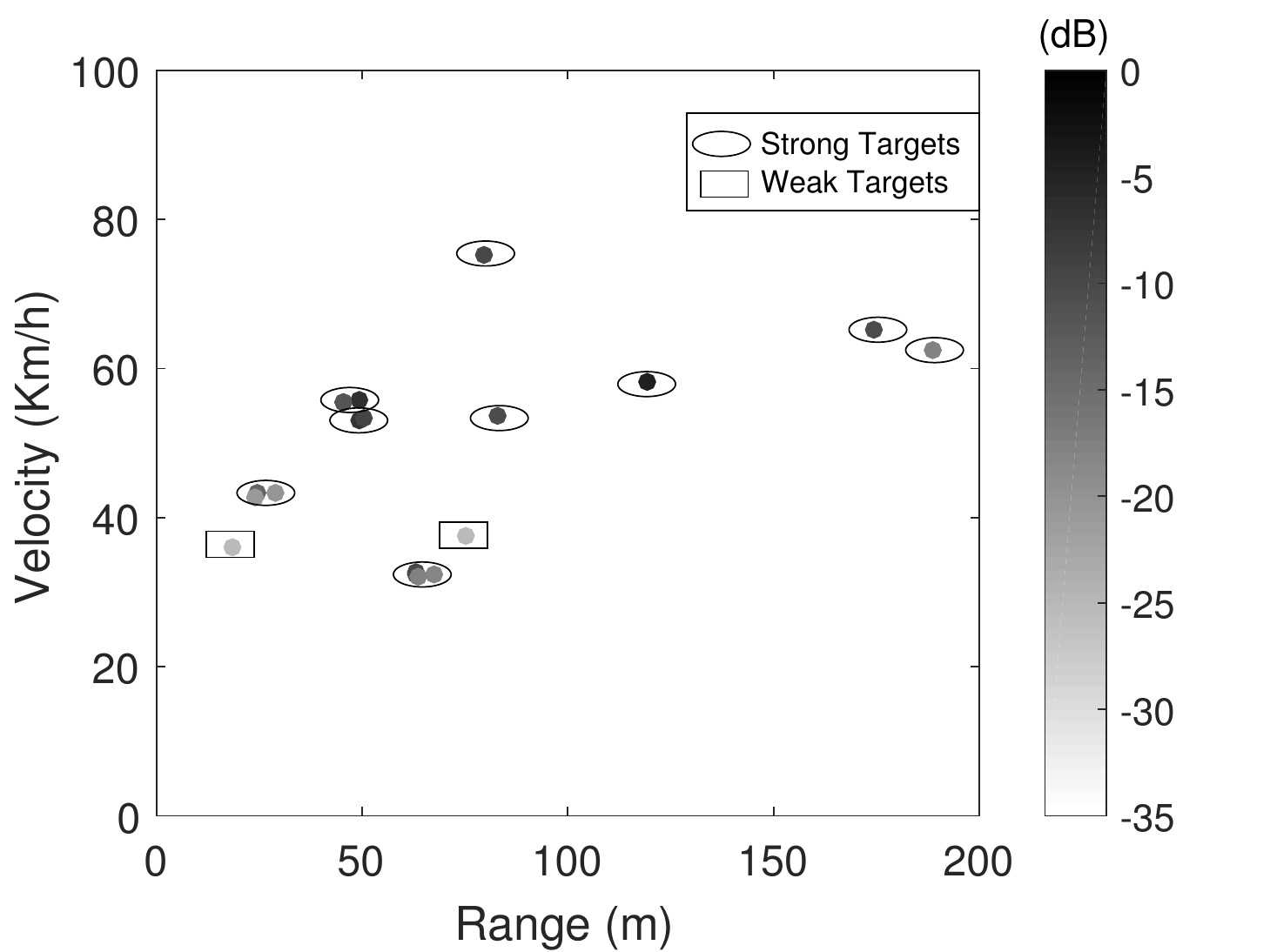}}
\caption{The range-Doppler image obtained with (a) 2D FFT and (b) conventional RELAX with BIC, both using original high-precision data, as well as with (c) 1bCLEAN and (d) 1bMMRELAX with 1bBIC, both using signed measurements. }
\label{RD}
\end{figure}

We plot the benchmark range-Doppler image obtained by applying 2-D FFT to the original high-precision data in Fig. \ref{RD}(a). There appears to be 9 strong targets and 2 weak targets in the scene of interest as well as background clutter. Fig. \ref{RD}(b) shows the range-Doppler image obtained from the high-precision data using the conventional RELAX algorithm with the conventional BIC \cite{PY05}. The estimate of the model order $K$ obtained via the conventional BIC is $\widehat K=81$. Inspecting the image in Fig. \ref{RD}(b) and comparing it with Fig. \ref{RD}(a), it can be seen that the conventional RELAX algorithm retrieves all the targets and quite a bit of the clutter from the high-precision measurements. The range-Doppler images obtained from the signed measurements using 1bCLEAN and 1bMMRELAX with 1bBIC are shown in Figs. \ref{RD}(c) and \ref{RD}(d), respectively. The corresponding model order estimates obtained are $\widehat K=7$ and $\widehat K=19$, respectively. We see that 1bMMRELAX provides excellent range-Doppler imaging performance with all the targets detected clearly, whereas 1bCLEAN misses quite a few targets. It is interesting that the clutter is absent in Figs. \ref{RD}(c) and \ref{RD}(d) while it is present in Fig. \ref{RD}(b). Note that we do not consider 1bRELAX for this 2-D application because it is computationally too demanding. Also, note that not every vehicle in the scene corresponds to a single scatterer (i.e., sinusoid) in the measured data. Consequently, multiple scatterers are usually needed to represent one vehicle in Figs. \ref{RD}(b)-\ref{RD}(d) for the experimentally measured data and therefore the estimated model order is not equal to the number of vehicles.

\section{Conclusions}
We have considered the problem of sinusoidal parameter estimation using signed measurements obtained with either fixed non-zero or time-varying thresholds. Making use of the MM technique, we have introduced the 1bMMRELAX algorithm with the main goal of improving the computational efficiency of the 1bRELAX algorithm. We have shown via multiple numerical examples that 1bMMRELAX can significantly reduce the computational complexity of 1bRELAX while maintaining its excellent estimation performance. We have also shown that 1bBIC performs well for model order determination when used with 1bMMRELAX. Furthermore, we have presented examples showing that both fixed non-zero and time-varying thresholds can be used to obtain accurate sinusoidal parameter and order estimates from one-bit measurements. Finally, experimental results have been presented to show that 1bMMRELAX with 1bBIC can be a useful technique for range-Doppler imaging in automotive radar applications.


%





\ifCLASSOPTIONcaptionsoff
  \newpage
\fi



\bibliographystyle{IEEEtran}
\bibliography{bare_jrnl_c}
\end{document}